\documentclass[british]{article}
\usepackage[left=1in,top=1in,right=1in,bottom=1in,nohead,paperwidth=8.5in, paperheight=11in]{geometry} 
\usepackage{soul}
\usepackage{amsmath}
\usepackage{caption}
\usepackage{stmaryrd}
\usepackage{natbib} 
\usepackage{float} 
\usepackage{rotating}
\usepackage{multirow}
\usepackage{algorithm}
\usepackage[titletoc, title]{appendix}
\usepackage{adjustbox}
\usepackage{graphicx}
\usepackage{bbm}
\usepackage{color}
\usepackage[table]{xcolor}

\graphicspath{{./Figures/}}

\usepackage[left=1in,top=1in,right=1in,bottom=1in,nohead,paperwidth=8.5in, paperheight=11in]{geometry} 
\usepackage{soul}
\usepackage{amsmath}
\usepackage{caption}
\usepackage{stmaryrd}
\usepackage{natbib} 
\usepackage{float} 
\usepackage{rotating}
\usepackage{multirow}
\usepackage{algorithm}
\usepackage[titletoc, title]{appendix}
\usepackage{adjustbox}
\usepackage{graphicx}
\usepackage{bbm}
\usepackage{color}
\usepackage[table]{xcolor} 
\usepackage{setspace}
\usepackage{subcaption}
\usepackage{lscape}

\graphicspath{{./Figures/}}

\begin{document} 

\begin{titlepage}
\title{\huge{\vspace{-2.5em}Emotions in Macroeconomic News and their Impact on \\ the European Bond Market}}

\author{Sergio Consoli\thanks{European Commission, Joint Research Centre, Directorate A - Strategy, Work Programme and Resources, Scientific Development Unit, Via E. Fermi 2749. I-21027 Ispra (VA), Italy [E-mail: sergio.CONSOLI@ec.europa.eu].} \and Luca Tiozzo Pezzoli\thanks{Corresponding author: European Commission, Joint Research Centre, Directorate A - Strategy, Work Programme and Resources, Scientific Development Unit, Via E. Fermi 2749. I-21027 Ispra (VA), Italy [E-mail: luca.TIOZZO-PEZZOLI@ec.europa.eu]} \and Elisa Tosetti\thanks{Cà Foscari University, Italy [E-mail: elisa.tosetti@unive.it]. 
}}


\vspace{-1.0em}
\date{This version: March, 2021.}
\maketitle

\footnotesize{

\begin{center}
{\large\bf Summary}
\end{center}


We show how emotions extracted from macroeconomic news can be used to explain and forecast future behaviour of sovereign bond yield spreads in Italy and Spain. We use a big, open-source, database known as Global Database of Events, Language and Tone to construct emotion indicators of bond market affective states. We find that negative emotions extracted from news improve the forecasting power of government yield spread models during distressed periods even after controlling for the number of negative words present in the text. In addition, stronger negative emotions, such as panic, reveal useful information for predicting changes in spread at the short-term horizon, while milder emotions, such as distress, are useful at longer time horizons. Emotions generated by the Italian political turmoil propagate to the Spanish news affecting this neighbourhood market.

\vspace{2.5em}

\vspace{0.5em}

 \vspace{17.5em}

\noindent \textit{The views expressed are purely those of the writers and may not in any circumstance be regarded as stating an official position of the European Commission. The authors declare that there is no conflict of interest that could be perceived as prejudicing of the research reported. This research did not receive any specific grant from funding agencies in the public, commercial, or non-for-profit sectors.} \vspace{0.5em}


}\normalsize

\end{titlepage}
\maketitle

\section{Introduction}

The turbulence in government yield spreads observed since the intensification of the financial crisis in 2009 in countries within the Euro area has originated an intense debate about the drivers of the sovereign bond market. 
Traditionally, factors such as public debt sustainability, liquidity of sovereign bonds, and global risk aversion as well as other macroeconomic variables like countries’ GDP growth and inflation have been indicated as important determinants of government yield spreads. Recently, an emerging literature having its roots in behavioural finance points at human perception, instinct and sentiment of investors as important elements that may guide their judgement and decision making, ultimately impacting their investment decisions (\cite{Blommestein2012}). To account for these factors, recent studies propose to measure sentiment from pieces of text such as news, blogs and other forms of written communication and use it to model and predict developments in the financial market (\cite{Tetlock_2007}, \cite{Garcia2013}). News often contain unanticipated information about the state of the economy in the form of description of the state of markets, comments on their evolution or about possible interventions in monetary policy. 
Through the lenses of news articles, market participants learn about recent economic events and trends, and adjust their perception and expectations on the dynamics of financial markets. A number of studies extract sentiment indicators from news and use it to model and forecast sovereign bond spreads (see, for example, \cite{Liu2014}, \cite{Apergis2015}, and \cite{Bernal_Gnabo_Guilmin_2016}). These works calculate their sentiment measures by looking at how many positive and negative words can be found in the text according to a predefined lexicon, such as the \cite{Loughran_McDonald_2011} word lists. Empirical findings provide evidence that news' sentiment conveys useful additional information over quantitative financial data for predicting government yield spreads. 
\newline

In this paper we extend this line of inquiry by focusing on \textit{emotions} extracted from macroeconomic news and their role in anticipating movements in sovereign bond spreads. Emotions, or affective states, are responses to interpretation and evaluation of events and hence reveal useful information about investors's behaviour (\cite{Ackert2003}). Different news for the same economic fact may express varying emotional states, depending on their interpretation of events and situations, inducing different reactions on the readers and leading them to different decisions. 
While commonly used sentiment measures are generally built from a binary classification (negative versus positive words), emotion extraction points at recognising the multivariate emotional content expressed by the text. As also stated by several works within the psychological literature, the tone and intensity of emotional words affect the processing, comprehension and memorisation of information (\cite{Megalakaki2019}). In this paper we argue that affective states such as distress, or fear, extracted from macroeconomic news may better capture, relative to simple univariate sentiment indicators used in previous studies, elements linked to human perception and mood that influence the behaviour and decision making of investors. 
\newline

We compute news-based emotion indicators from macroeconomic news for two countries, Italy and Spain, and employ them to forecast daily changes in 10-years government bond spreads dynamics over the period March 2015 until end of August 2019. 
We use a novel, real-time, open-source news database, known as Global Knowledge Graph (GKG) provided by the Global Dataset of Events, Language and Tone (GDELT) data platform (\cite{Leetaru_2013gDELT}).
We extract the emotional content of macroeconomic news by applying the WordNet-Affect dictionary\footnote{See http://wndomains.fbk.eu/wnaffect.html (last access 27 September 2020)} by \cite{Strapparava2004} and \cite{Strapparava2004a}. 
By adopting such classification scheme we are able to measure the presence of specific emotions in the selected macroeconomic news and use them as predictors of market investor's expectations and behaviour. We focus on \textit{negative} emotions, conveying varying intensities of fear and linked to the different affective states of investors within part of the so-called cycle of market emotions (\cite{Taffler2018}).\newline

We exploit GDELT's locations extraction algorithm to determine if an article from a national outlet is predominantly talking about domestic events or foreign ones. The aim is to distinguish emotions caused by domestic facts from those elicited by foreign events and evaluate their impact on the national yield spread, thus allowing us to investigate contagion effects across economies. 
We adopt a quantile regressions approach (\cite{Koenker2011}), and carry an out-of-sample analysis focusing on the 95th percentile of sovereign bond spread. One motivation for using quantile regression is that we expect emotions to affect the spread particularly during distressed periods, potentially impacting on the right tail of our variable of interest. Another reason is that quantile regressions is a useful tool for dealing with non-linearity and parameter instability that are present in our data given the high political turbolent period of 2018.\newline

We evaluate the predictive performance of our emotion-augmented model with respect to a benchmark model where we include the traditional determinants of government bond spreads as well as the \cite{Loughran_McDonald_2011} (LM) overall measure of negativity. The rationale for including the LM indicator in our regression specification is to test whether there is any statistically significant incremental information from including emotions relative to an overall measure of negativity. We look specifically at the length of the time effect of the different emotions with the intent to verify whether emotions intensity is linked to the forecasting horizon of government spreads predictions. 
We use the Fluctuation test by \cite{Giacomini2010} to evaluate the predictive power of our emotional measures.\bigskip \newline

Our results show that augmenting quantile regression with selected daily news emotions improves significantly the in-sample prediction of the spread right tail relative to the benchmark model during distressed periods for both Italy and Spain. While the Italian spread is mostly affected by emotions triggered by domestic facts, variations in the Spanish spread are generally anticipated by emotions elicited by foreign events.
We also find that negative emotions extracted from news improve the forecasting power of government yield spread models during turbolent periods, although this is only valid for Italy. One interesting result is that the time length of news effects on the spread seems linked to the intensity of the emotion, at least for Italy. In particular, relatively lower intensity emotions can help predict fluctuations in the spread as long as a week after, while relatively higher intensity emotions are able to predict one-day ahead changes in the spread. This could be explained by the fact that a rise in relatively lower intensity emotions tend to precede an increase in relatively higher intensity emotions (\cite{Taffler2018}).\newline

The remainder of the paper is structured as follows. Section \ref{S1} reviews existing literature on this topic. Section \ref{S2} introduces the data, while Section \ref{s_emotions} is devoted to the construction of emotion indicators. Section \ref{S4} briefly describes the methods and Section \ref{S5} comments on empirical results. Finally, Section \ref{S6} concludes with some concluding remarks.

\section{Background literature}
\label{S1}

\subsection{Extracting sentiment from economic text}

In the last two decades, a growing body of literature has been studying the role of textual sentiment extracted from economic news on stock returns and trading volume. A critical aspect of these studies is the choice of the method used for extracting a sentiment score from the text of the news analyzed. One simple approach consists of searching specific keywords of interest within the text of the article to devise news' sentiment. In the influential paper by \cite{Bloom_Baker_Scott_2016}, the authors searched the terms: ``economic'', ``policy'', ``uncertainty'', in a set of newspapers articles, in order to construct an indicator of policy-related economic uncertainty, known as EPU\footnote{EPU index: https://www.policyuncertainty.com/ (last access 27 September 2020)}, for the US economy. The authors found that a rise in the EPU index for the US predicts a decline in investment, output, and employment, while it raises US stock price volatility. The EPU index has been used extensively as a proxy for economic uncertainty in various economic and financial applications (see, among others, \cite{Chen2017} and \cite{Bernal_Gnabo_Guilmin_2016}). 
\newline

A number of works calculate the sentiment of news by adopting \textit{lexicon-based approaches}. These methods require the use of a predefined dictionary, or lexicon, of words carrying on a certain positive or negative sentiment polarity, and involve counting dictionary words found in the article text to compute an overall sentiment score for the entire article. A commonly used source for sentiment word classification is the ``Harvard IV-4 dictionary''\footnote{For the Harvard IV-4 dictionary, see: http://www.wjh.harvard.edu/~inquirer/homecat.htm (last access 27 September 2020)}, listing negative and positive words categorized from a general or psychological perspective. For example, \cite{Tetlock_Tsechansky_Macskassy_2008} showed that the fraction of negative words from the Harvard IV-4 negative word list found in firm-specific news stories helps forecasting firms’ stock prices. \cite{Loughran_McDonald_2011} emphasized that the Harvard IV-4 word list might not be suitable for finance and accounting applications, given that many terms that the list classifies as positive or negative from a general or psychological point of view might not have the same connotation within a financial context. Hence, the authors extended this list to include negative, positive and uncertain terms categorized from a business and finance perspective by parsing 10-K reports and creating the so-called ``Loughran and McDonald Sentiment Word Lists'' dictionary\footnote{For the ``Loughran and McDonald Sentiment Word Lists'' dictionary, see: https://sraf.ndanalysis/resources/(last access 27 September 2020)}. 
\cite{Garcia2013} computed the fraction of Loughran and McDonald's positive and negative words extracted from New York Times articles to predict variations in the Dow Jones stock market index. The author showed that predictability is stronger during recession since investors are more sensitive to negative news stories during periods of economic troubles (see also \cite{Goetzmann2016} on this point). 
\newline

Recently, more sophisticated techniques from the Natural Language Processing (NLP) literature have been employed to extract information from news and analyze its impact on financial and economic variables. We refer to \cite{Gentzkow2019} for a recent overview on the subject. Some studies have adopted Latent Dirichlet Allocation to classify articles in topics and calculate simple measures of sentiment based on the topic classification to predict economic activity (\cite{thorsrud2016}, \cite{Thorsrud_2018} and \cite{shapiro2018}).
\cite{Dridi2018} developed a fine-grained sentiment analysis algorithm to predict optimistic and pessimistic market moods from micro-blog and Twitter messages, where with the term ``fine-grained'' (\cite{Reforgiato2015211}) is intended that the detected sentiment polarity is given by means of a continuous value within a certain range (e.g. $[-1,+1]$). 
\cite{Barbaglia2019} proposed an unsupervised, rule-based procedure that exploits the semantic structure of pieces of news to calculate the sentiment for a given economic concept mentioned in the news text.
The range of tools available from the most recent NLP literature is very wide and its use in the area of economics and finance is still largely unexplored, with significant potentials for further study.\bigskip
\newline

Rather than calculating an univariate sentiment score, a recent strand of research attempts at detecting emotions from text as well as other media sources, looking at their impact on economic outcomes. 

In particular, \cite{MAYEW2012} measured managers' emotional state by analyzing their earnings conference call audio files in 2007. The authors measured the positive (e.g., happiness, excitement, and enjoyment) and negative (e.g., fear, tension, and anxiety) dimensions of a manager’s affective or emotional state. Even after controlling for the number of negative words present in the speech, as measured by the \cite{Loughran_McDonald_2011} negative words list\footnote{Available at: https://sraf.nd.edu/textual-analysis/ (last access 27 September 2020)}, \cite{MAYEW2012} showed that higher levels of negative affect expressed by the voice of managers conveys bad news about future firm performance. We refer to \cite{ALLEE2015} for further work on this. \cite{Yuan2018} developed a method to extract the distribution of eight public's emotions from textual postings on online social media to supplement common financial indicators (e.g., return-on-assets) for predicting corporate credit ratings. Using real-world data crawled from Twitter, the authors showed that the extracted emotion dimensions enhance the prediction of corporate credit ratings by using an ensemble learning model with random forest as the basis classifier. \cite{Griffith2019} explored the interaction between media content, market returns and volatility. They selected four measures of investor emotions that reflect both pessimism and optimism of small investors (i.e. ``fear'', ``gloom'', ``joy'', and ``stress'') from Thomson Reuters MarketPsych\footnote{Thomson Reuters MarketPsych Indeces: https://www.marketpsych.com/ (last access 27 September 2020)}, a propriety set of investors' emotional measures calculated from news and social media. The authors explored the ability of these emotional measures to predict both level and change in market returns.

\subsection{News sentiment and government yield spreads}\label{s1_1}

Existing studies on government bond yields modelling generally point at three important drivers for bond interest rates spreads, namely credit risk, liquidity risk and global risk aversion (see \cite{Codogno2003}, \cite{attinasi2009}, \cite{Schwartz_2018} among others). In particular, credit risk refers to the probability of a country to default on its debt, and is often approximated by Credit Default Swaps (CDS) spreads (see \cite{Baber_Brandt_Kavajecz_2009} and \cite{Favero_2013}), although some other authors use the daily return of the domestic stock market of the country (see \cite{Afonso_Fuceri_Gomes_2012}, \cite{Olivera_Curto_Nunes_2012} and \cite{Bernal_Gnabo_Guilmin_2016}, among others). Liquidity risk captures the possibility of capital losses due to early liquidation or significant price reductions resulting from a small number of transactions. This variable is usually approximated using bid-ask spreads, transaction volumes and the share of a country’s debt within the global sovereign debt (\cite{Desantis_2012} and \cite{Favero_2013}, among others). Finally, global risk aversion measures the level of perceived financial risk and its unit price. This variable is empirically approximated using indexes of stock market implied volatility (\cite{Kilponen2015}, \cite{Baber_Brandt_Kavajecz_2009} and \cite{Bernal_Gnabo_Guilmin_2016}, among others). 
\newline

More recently, a new strand of literature from the field of behavioural finance (\cite{Shiller2003}) have tried to incorporate in their models for government yield spread factors related to human perception, mood and emotional reaction. These may guide judgement and risk aversion of investors, particularly during periods of crisis, ultimately affecting their decision making (\cite{Ackert2003}, \cite{Fenton2011}). 
Recently, a number of studies have focused on sentiment extracted from pieces of text to account for mood and emotional states of investors.
 \cite{Mohl2013} scanned news agency reports for statements about ``restructuring'', ``bailout'' and the ``European Financial Stability Mechanism'' in the years 2010-2011. They found that the tone of these statements impacted negatively bond spreads in five peripheral Euro area countries. Using news media releases over the period from 2009 to 2011, \cite{Gade2013} studied the impact of political communication on the sovereign bond spreads of three peripheral European countries. The authors developed an algorithm based on keyword search and basic semantic analysis to quantify whether political communication has a positive or negative tone, founding that this has a contemporaneous effect on yield spreads. \cite{BEETSMA2013} extracted information from daily morning news briefings of European media to construct a set of risk variables based on the volume of news released in a country in a given date. According to the study, an increase in the volume of bad news amplify the turbulence in the market with a consequent raise in the country-specific interest rate and spillover effects to other economies. The volume of news and their pessimism calculated using the Loughran and McDonald negative word list has been examined also by \cite{Liu2014} for the GIIPS countries over the period 2009 to 2012. The author found that higher media pessimism coupled with larger volume of news yield spreads move upwards, causing price to fall. \cite{Bernal_Gnabo_Guilmin_2016} exploited the EPU index developed by \cite{Bloom_Baker_Scott_2016} to study the impact of economic policy uncertainty on risk spillovers within the Euro zone. The authors showed that an increased level of uncertainty in one economy raises the risk that shocks in that country would affect the entire Euro area. \cite{Apergis2015} explored the forecasting performance of sentiment extracted from newswire messages for Credit Default Swaps (CDS) for five European countries. The author showed the better forecasting performance of an ARIMA model for CDS augmented with news sentiment relative to a pure time series model, pointing at the importance of news sentiment for better risk profiling of a country. Similar results have been obtained by \cite{Apergis2016}, using a panel ARDL and asymmetric conditional volatility modelling methods.
\newline

Despite the different approaches and frameworks, there is general consensus in the studies reviewed above that sentiment and emotions extracted from news convey valuable information in addition to financial variables for explaining and predicting stock and bond markets dynamics. Most of studies extract sentiment from news by searching specific keywords in the text, or by looking at how many positive and negative words can be found in the text according to a predefined lexicon. In addition, most of the studies reviewed above carry in-sample analyses, while the power of textual sentiment for forecasting future movements of bond markets in out-of-sample studies has been largely overlooked and will be the object of this work. 

\section{Data}
 \label{S2}

\subsection{Yield spread}

We extracted data from Bloomberg on the Italian and the Spanish 10-year government bond daily yield spreads over the period 2 March 2015 to 31 August 2019. The sovereign bond yield spread for a country is defined as 10-year government bond index minus the German counterpart, where German bonds are considered as the risk-free asset for Europe (see, for example, \cite{Afonso_Arghyrou_Kontonikas_2015}). We take the most recently issued bonds, i.e. on-the-run bonds, given that these are the most traded bonds with the smallest liquidity premium.
\newline

\begin{figure}[H]
\centering
\caption{10-year Italian (black) and Spanish (red) government bond yield spreads with ID of events.} 
\includegraphics[scale=0.3]{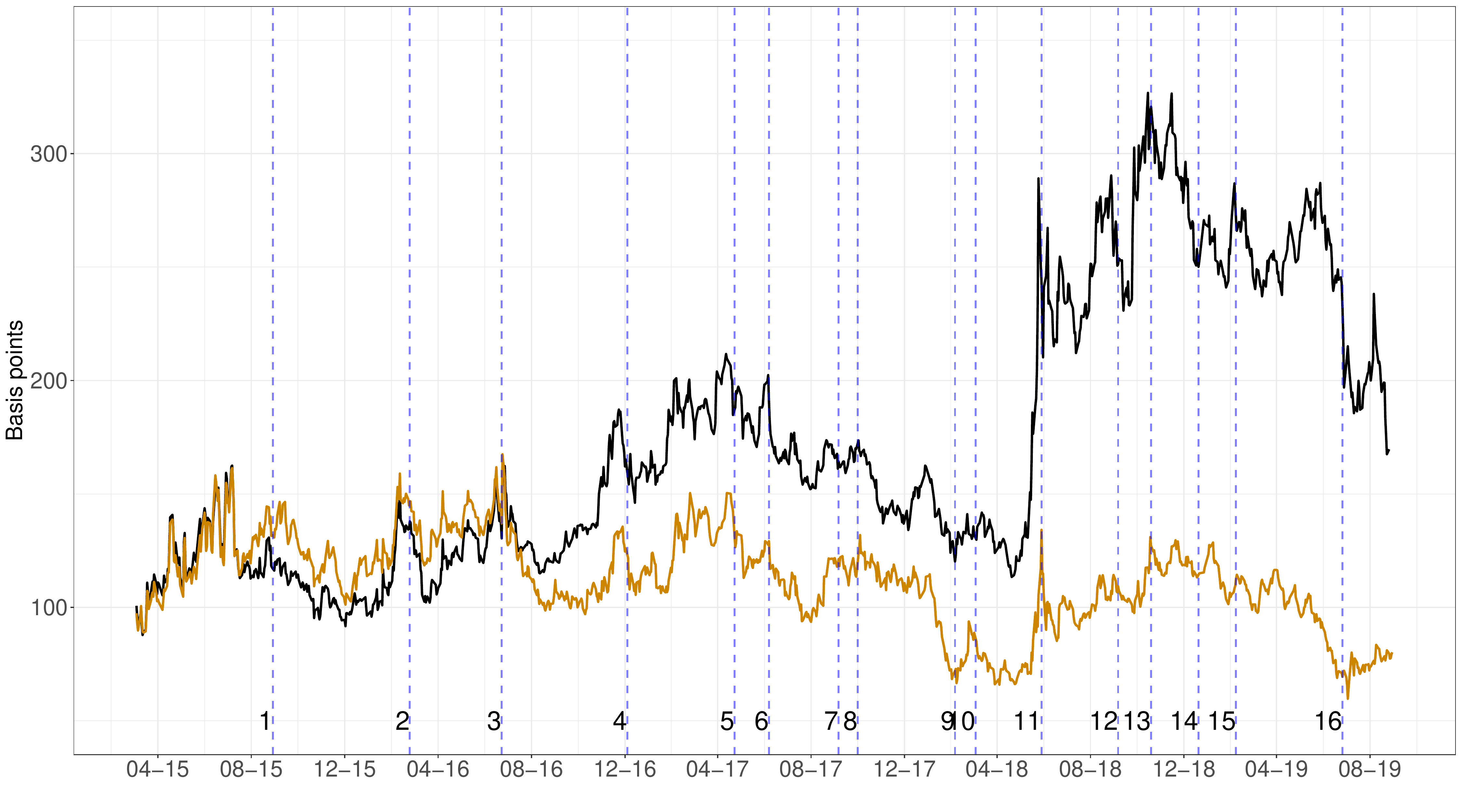}
\label{fig1}
\end{figure}

The temporal dynamics of the yield spread for the two countries over the sample period are plotted in Figure \ref{fig1}, with the vertical dashed lines indicating the timing of some important, stressing, events that are listed in Table \ref{tab_events}. The behaviour of the Italian and Spanish spreads moved closely till late June 2016 when they started diverging in occasion of the Brexit referendum and Spanish elections, when a wave of uncertainty hit investors, that moved away from the riskier Italian market.
At the end of May 2018, a period of high political turmoil started in both countries. In Spain a motion of no confidence was held against the prime minister. In Italy the spread sharply rose, passing from around 100 basis points and reaching a peak of over 250 basis point on the 29th of May 2018, and remained well above 200 basis points afterwards. The Italian spread lowered in June when the government was formed, but it rose again reaching 350 base points on the 19th of October, followed by another peak in November when investors started worrying about deficit spending engagements of the new government and possible conflicts with the European fiscal rules. In the same period, a wave of anxiety propagated in Europe, causing an increase in borrowing costs especially in countries from Southern Europe. 
In 2019, the Italian and the Spanish spreads generally declined, although some events hit the Italian economy, such as the EU negative economic outlook towards Italy and the European parliament elections. These facts contributed to a temporary increase of Italian rates in February, May and August. Although, the Italian interest rates jumped much higher than their Spanish counterparts and risk spillovers seemed to be contained\footnote{This phenomenon was also remarked by the president of the European Central Bank, Mario Draghi, during the ECB Governor Council meeting in Riga in June 15, 2018}, Southern European economies appeared to be closely related to the evolution of the political situation in Italy. The timing of these events are indicated in Figure \ref{fig1} with blue vertical dotted lines. The events can be interpreted as stressing events, namely a set of economic and political events, both domestic and international, that are likely to have triggered a reaction on investors and hence on the Italian and Spanish spreads. 
\newline

\begin{figure}[H]
\footnotesize
\caption{Government yield spread (in first differences), credit risk (market's returns), liquidity risk (in first differences) for Italy.} 
\begin{center}
{\includegraphics[scale=0.26]{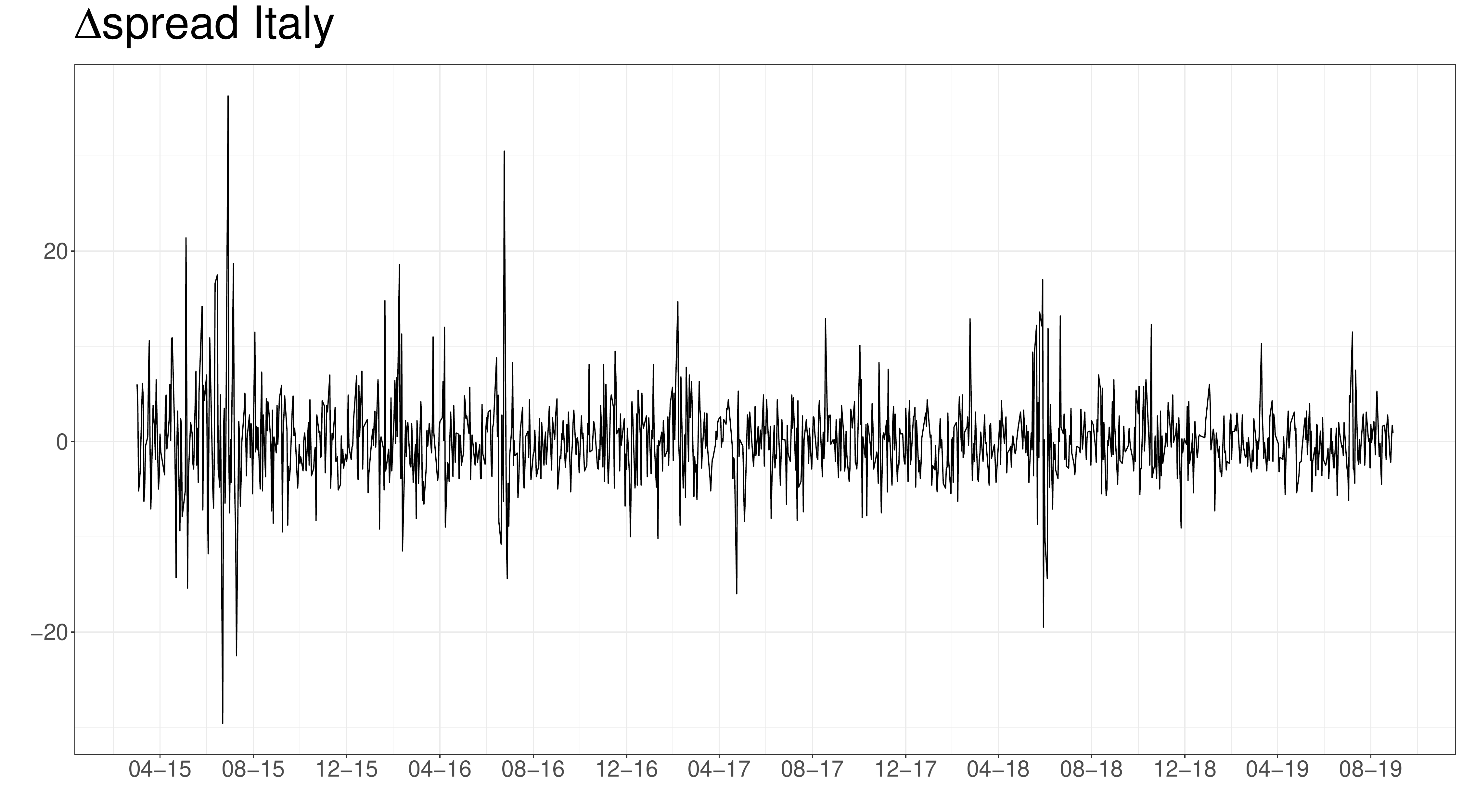}}
\end{center}
\begin{center}
{\includegraphics[scale=0.26]{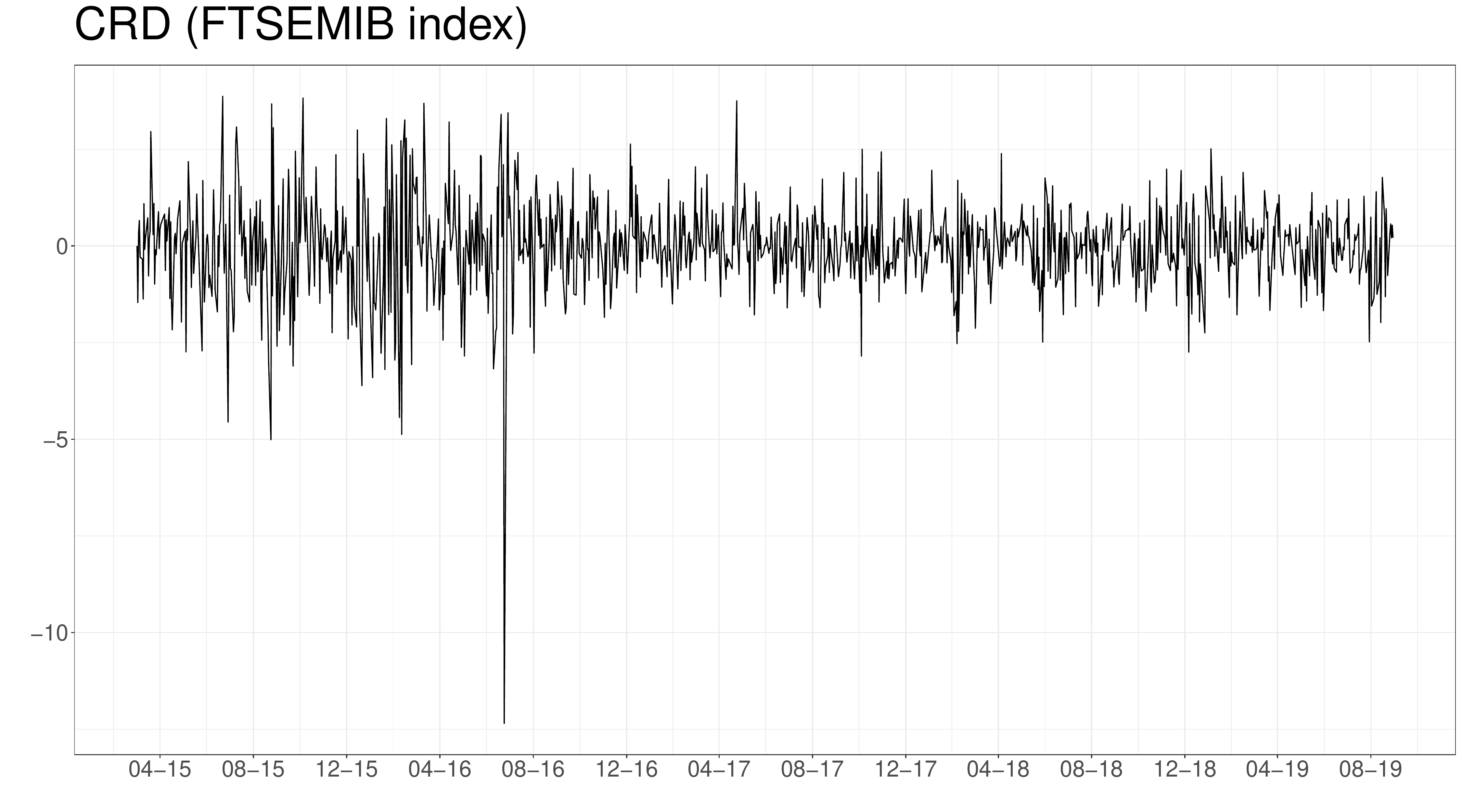}}
\end{center}
\begin{center}
{\includegraphics[scale=0.26]{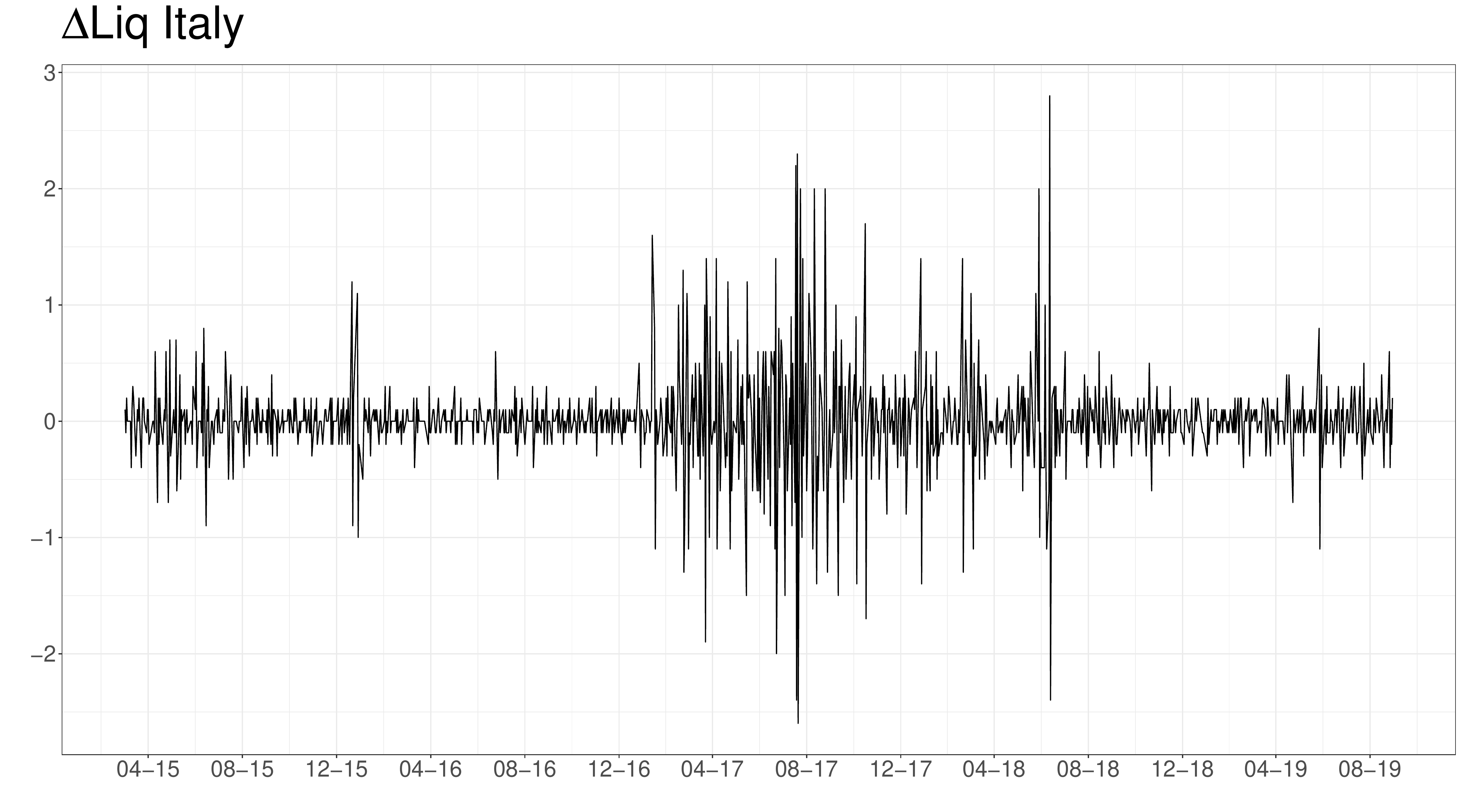}}
\end{center}
\label{fig_spread1}
\end{figure}

\begin{figure}[H]
\footnotesize
\caption{Government yield spread (in first differences), credit risk (market's returns), liquidity risk (in first differences) for Spain.} 
\begin{center}
{\includegraphics[scale=0.26]{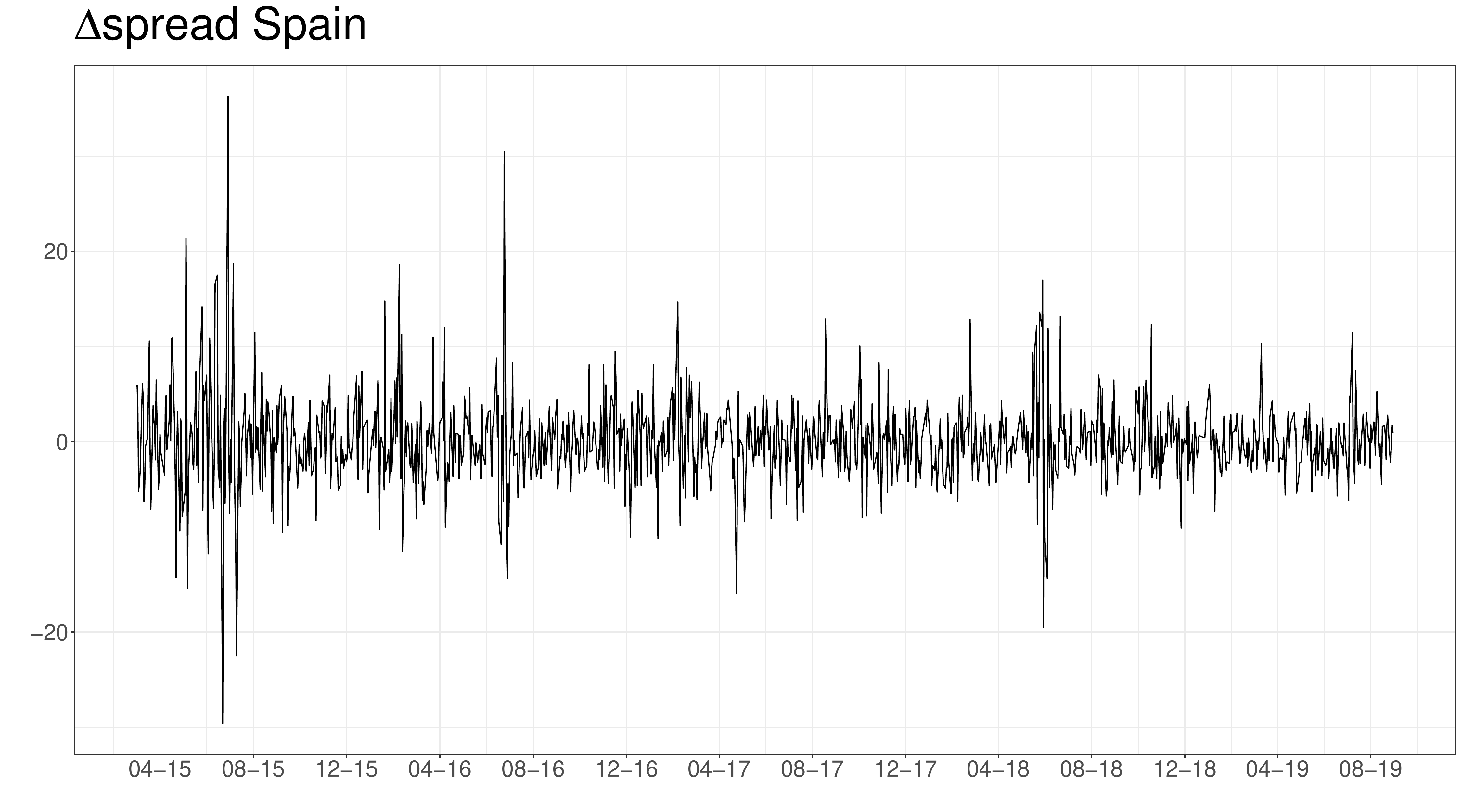}}
\end{center}
\begin{center}
{\includegraphics[scale=0.26]{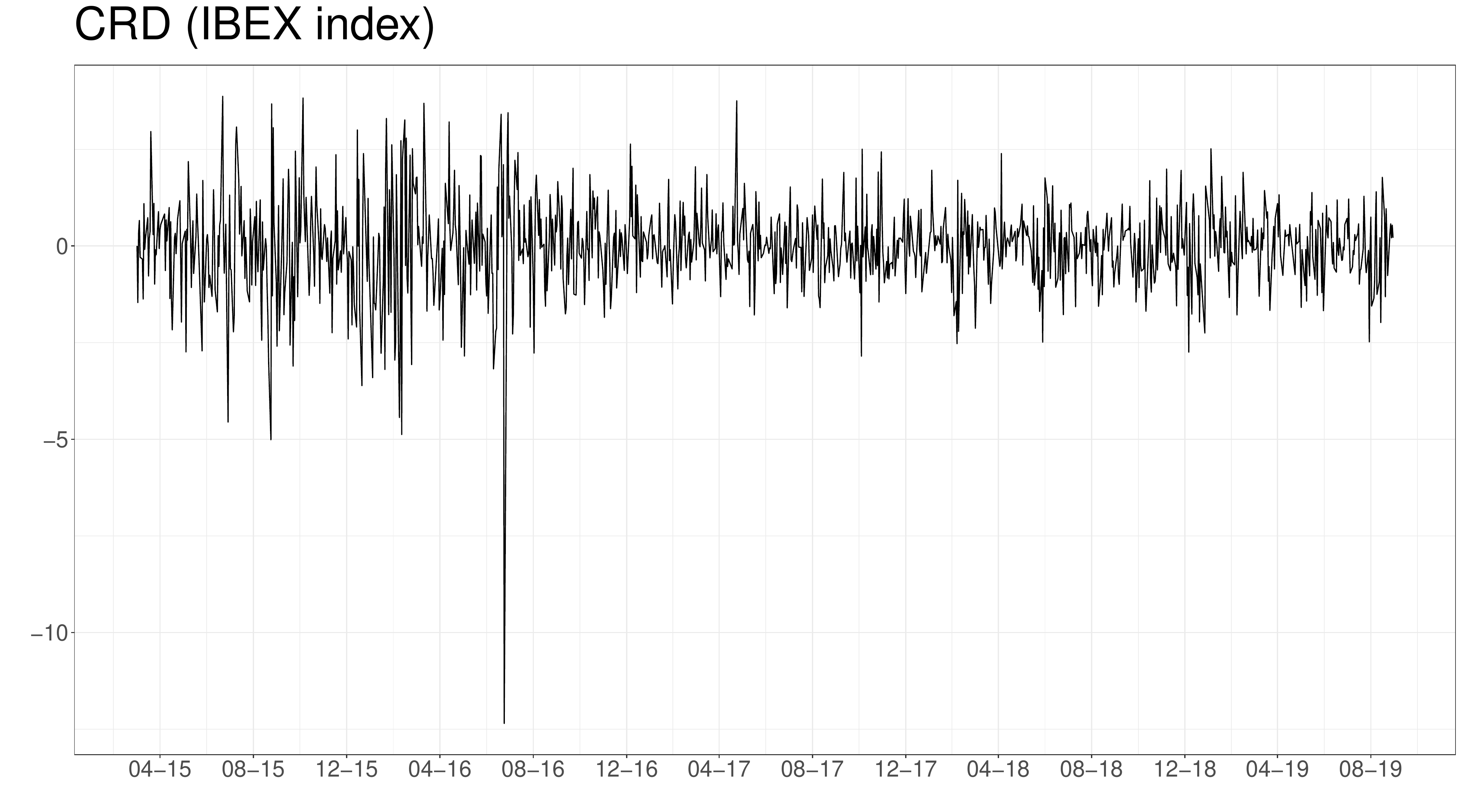}}
\end{center}
\begin{center}
{\includegraphics[scale=0.26]{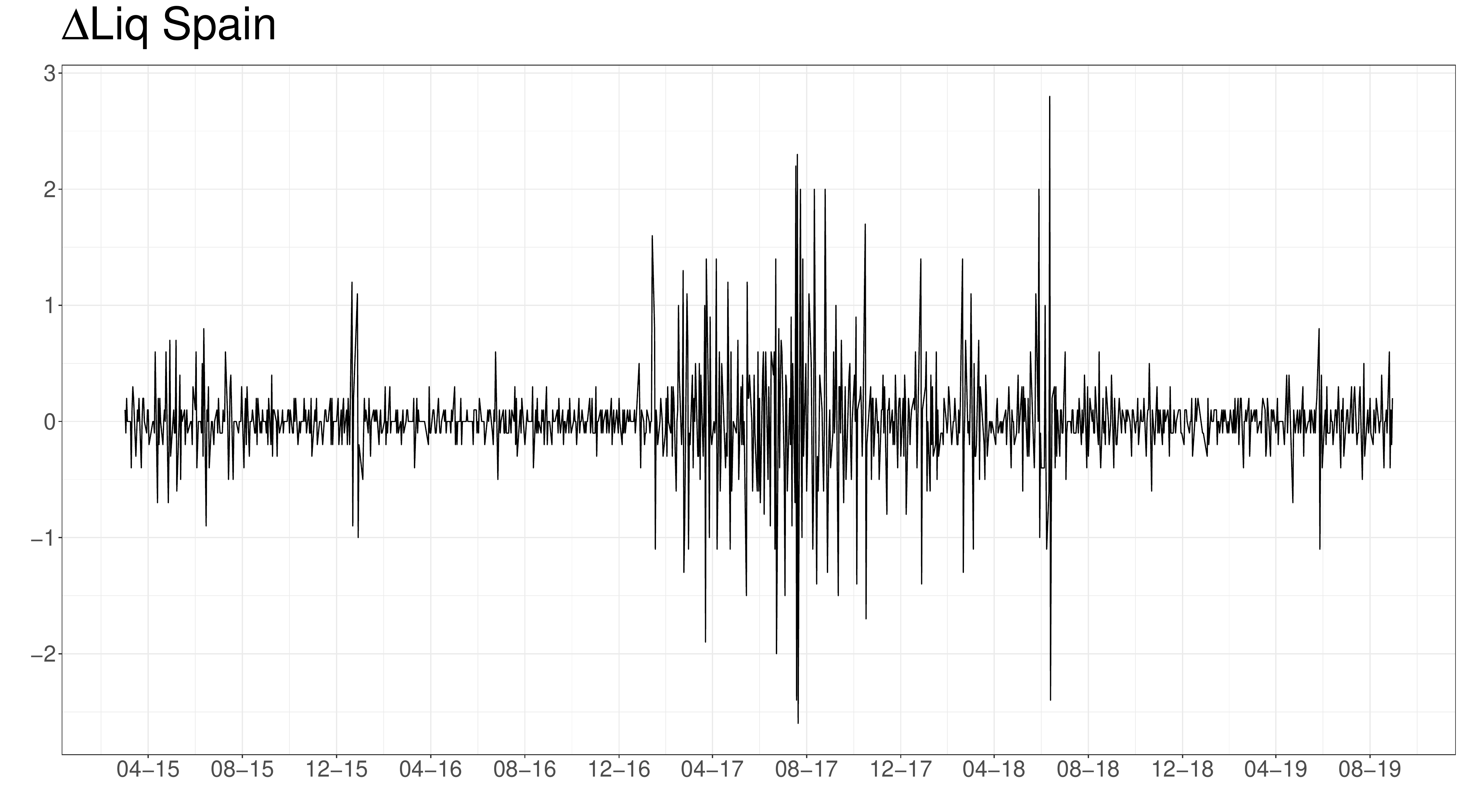}}
\end{center}
\label{fig_spread2}
\end{figure}

\begin{figure}[H]
\footnotesize
\caption{European Implied Volatility Index (VSTOXX) (in first differences).} 
\begin{center}
\includegraphics[scale=0.28]{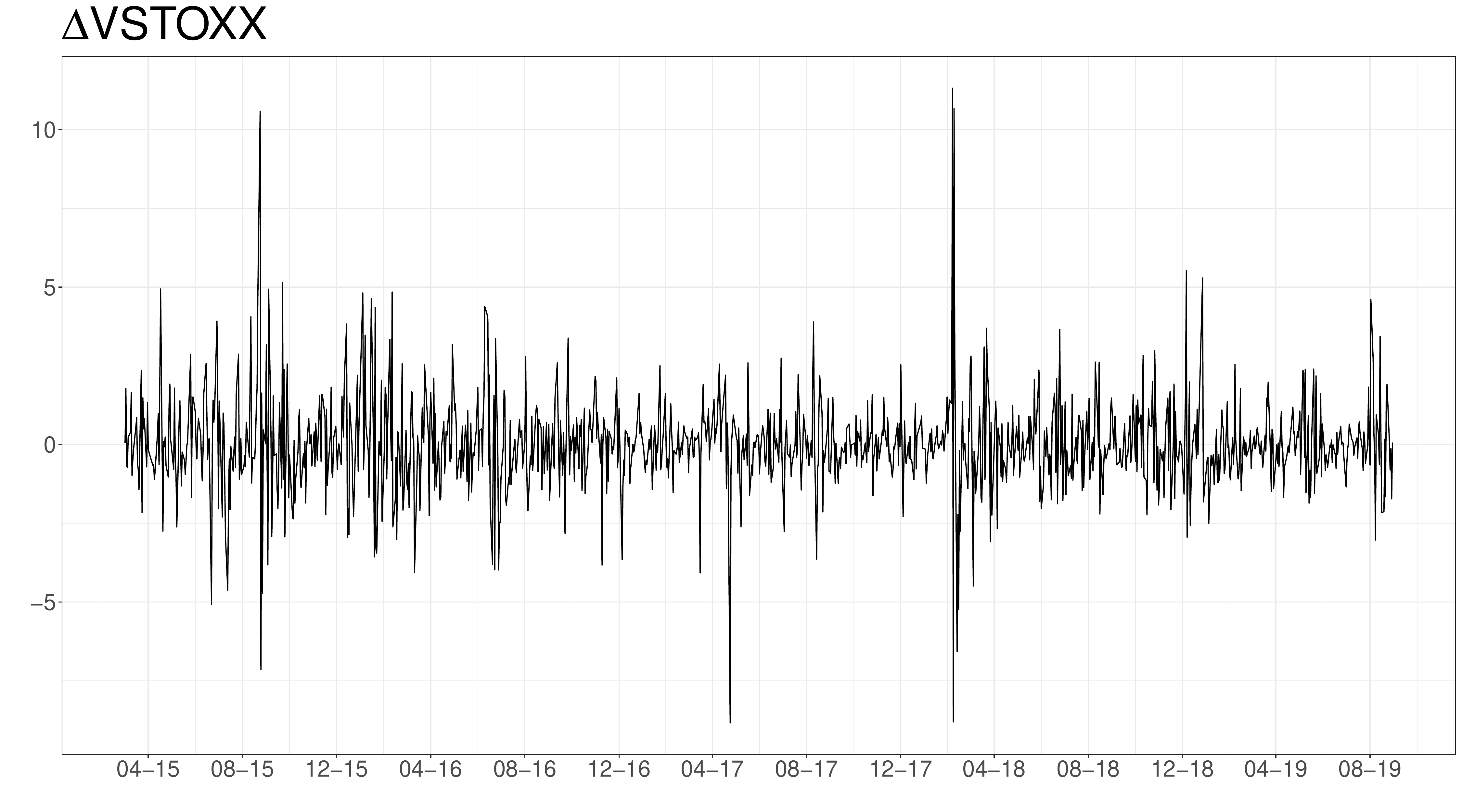}
\end{center}
\label{fig_vstoxx}
\end{figure}

In our model for forecasting sovereign bond yields spreads, we also include a set of variables that are traditionally included as determinants of sovereign bond yields spreads, namely credit risk, liquidity risk and global risk aversion (see Section \ref{s1_1}). 
In particular, we use daily returns of FTSEMIB and IBEX for Italy and Spain, respectively as proxies for credit risk, we measure liquidity risk by taking the bid-ask spread of a country 10-year government bond yield, and we approximate global risk aversion by the European Implied volatility index ($VSTOXX$). These variables have been collected from Bloomberg. Summary statistics by year and by country for these variables are reported in Table \ref{tab1}. We note that, since $VSTOXX$ is calculated at European level, we only report statistics for this variable for Italy as it is identical for Spain.
Figures \ref{fig_spread1}-\ref{fig_spread2} show the temporal evolution of spread (expressed in first differences), domestic market returns and liquidity risk (expressed in first differences) for Italy and Spain, respectively. In Figure \ref{fig_vstoxx} we have also plotted the European investor's risk aversion (expressed in first differences). It is interesting to observe that in both countries, credit risk reacts to some of the events listed in Table \ref{tab_events} and having an impact on government spread across-countries, such as the Greek crisis, the EU stress test, or the Brexit referendum. By building our news-based indicators we aim at explaining variations in the government yield spread due to national facts that are not captured by conventional determinants.

\subsection{News data}

GDELT (Global Database of Events, Language and Tone) is an open, big data platform of meta-information extracted from broadcast, print, and web news collected at worldwide level and translated nearly in real-time into English from over 65 different languages (\cite{Leetaru_2013gDELT}).\footnote{https://www.gdeltproject.org/ (last access 27 September 2020)} 
It collects, translates into English, and processes news worldwide, and updates them on a dedicated web-platform every 15 minutes.\footnote{See http://data.gdeltproject.org/gdeltv2/lastupdate.txt for the English version, while http://data.gdeltproject.org/gdeltv2/lastupdate-translation.txt for the translated version (last access 27 September 2020).}
Three primary data streams are created, one codifying human activities around the world in over 300 categories, one recording people, places, organizations, millions of themes and thousands of emotions underlying those events and their interconnection, and one codifying the visual narratives of the world's news imagery. Extracted and processed information are stored into different databases, with the most comprehensive among these being the GDELT Global Knowledge Graph (GKG). GKG is a news-level data set, containing a rich and diverse array of information. Specifically, for each news GKG extracts information on people, locations and organizations mentioned in the news, it retrieves counts, quotes, images and themes present in the news using a number of popular topical taxonomies, such as the World Bank Topical Taxonomy (WB)\footnote{https://vocabulary.worldbank.org/taxonomy.html (last access 27 September 2020).}, or the GDELT built-in topical taxonomy. Finally, it computes a large number of emotional dimensions expressed by means of commonly used dictionaries, such as the Harvard IV-4 Psycho-social dictionary, the Loughran and McDonald word list, or the WordNet-Affect dictionary. Specifically, it extracts over 2,200 dimensions, known as Global Content Analysis Measures (GCAM). The output of such processing is updated on the GDELT website every fifteen minutes and is freely available to users by means of custom REST APIs. In terms of volume, GKG analyses over 88 million articles a year and more than 150,000 news outlets. Its dimension is around 8 TB, growing approximately 2 TB each year. To be able to process the huge amount of unstructured documents coming from GDELT, we built an ad-hoc Elasticsearch infrastructure (see \citet*{Gormley_Tong_2015} for more details). Elasticsearch is a popular and efficient document-store built on the Apache Lucene\footnote{https://lucene.apache.org/ (last access 24 March 2021)} search library, providing real-time search and analytics for different types of complex data structures. News data have been downloaded from the GDELT website, re-engineered and stored into our Elasticsearch platform. This has allowed us to efficiently store and index data in a way that supports fast search, data retrieval and processing via simple REST APIs.\newline


We have extracted news information from GKG from a set of around 20 newspapers for each of the examined countries, Italy and Spain, published over the sample period. The chosen newspapers include both generalist national newspapers with the widest circulation in that country, as well as specialized financial and economic outlets (Appendix A provides the list of newspapers considered in the analysis). Once collected the news data, we have mapped these to the relevant trading day. Specifically, we assign to a given trading day all the articles published during the opening hours of the bond market, namely between 9.00am and 17.30pm. Articles that have been published after the closure of the bond market or overnight are assigned to the following trading day.\footnote{Since the GKG operates on the UTC time, https://blog.gdeltproject.org/new-gkg-2-0-article-metadata-fields/ (last access 27 September 2020), we made a one-hour lag adjustment according to Italian and Spanish time zone.} Following \cite*{Garcia2013}, we assign the news published during weekends to Monday trading days, and omit articles published during holidays or in weekends preceding holidays. 
Once extracted news information from GKG, we have cleaned the data in various ways. First, to obtain a pool of news that are not too heterogeneous in length, we have retained only articles that are long at least 100 words.\footnote{Such cleaning operation implies dropping only a very small number of articles. For Italy the total number of articles without the 100 limit on the word count is 9,234  while with such limit is 9,119 (1.26$\%$  increment). 
For Spain we pass from 12,209 without the word limit to 12,203 (0.05$\%$ increment).} Given that we wish to measure emotions related to events concerning bond market investors, rather then event in general, we have exploited information from the World Bank Topical Taxonomy to understand the primary focus (theme) of each article and select the relevant news. Such taxonomy is a classification schema for describing the World Bank’s areas of expertise and knowledge domains representing the language used by domain experts. 
Hence, we have selected only articles such that the topics extracted by GDELT fall into one of the following WB areas of interest: \textit{Macroeconomic Vulnerability and Debt}, and \textit{Macroeconomic and Structural Policies}. In particular, we have retained news that contain in their text 
at least four keywords belonging to these themes. The aim is to select news that focus on topics relevant for the bond market, while excluding news that only briefly report macroeconomic, debt and structural policies issues.\footnote{We have done some robustness checks on the number of keywords from the WB themes required for including a news in the data set, by varying it from one to five. Results on the size and significance of the effects of news on the spread are quite stable when varying this parameter between 1 and 4, while when setting the threshold to 5 we lose some significance perhaps due to the drop in the number of selected articles. Results are available upon request from the authors.}\newline

A final filter that we applied to our news selection concerns the locations extracted from the text of the news. Given that we wish to analyze news in a country that predominantly talk about national events, we have only retained articles for which the main location was, respectively, Italy or Spain. To this end, we have used the \textit{locations} information in GKG to infer the main location mentioned in an article. Specifically, to obtain, for example, all articles that talk about events in Italy, we have only retained those articles that mention Italy more frequently than all other remaining mentioned countries. The same procedure has been followed for Spain. After this selection procedure we have obtained a data set of 9,119 articles for Italy, and 12,203 for Spain. In a separate analysis, we have also considered Italian and Spanish articles that mention Spain and Italy together in the same article, thus having an \textit{international} focus. This is done with the idea of capturing domestic emotions triggered by foreign events, possibly due to information contagion effects. To select such set of articles we have taken all Italian news with main location Italy or Spain or \textit{both}, and then all Spanish news with main location Spain or Italy or \textit{both}. By doing this we have selected a total of 9,458 articles for Italy, and 14,479 for Spain.
\newline

\begin{figure}[H]
\caption{Number of articles by country of publication speaking about national and both national and international events.} 
\vspace*{5mm}
\begin{center}
\subcaption{Italy. Articles about domestic events (black) and  both  national  and international events (red).}
\vspace*{5mm}
\centering
\includegraphics[scale=0.35]{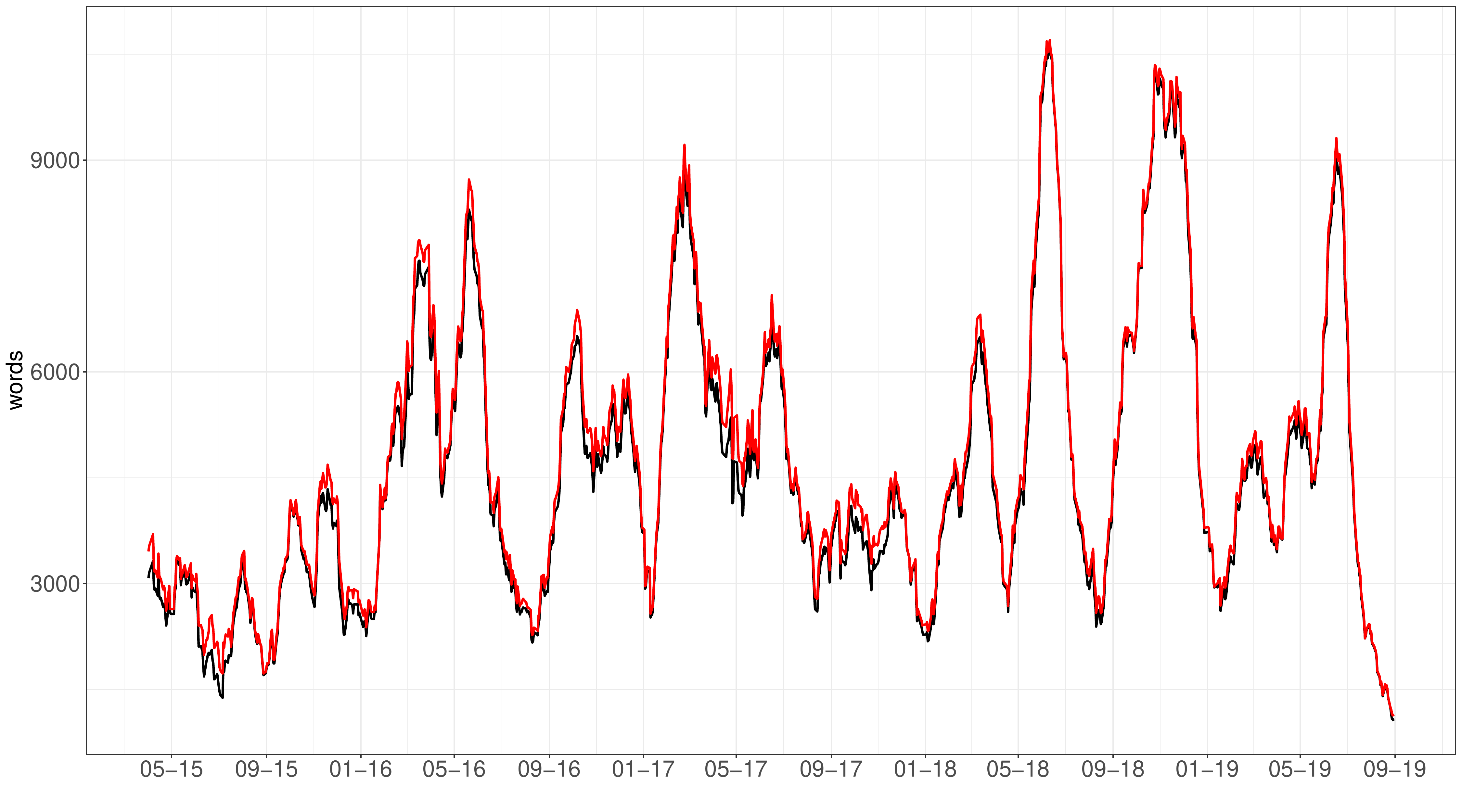}
\end{center}
\begin{center}
\vspace*{5mm}
\subcaption{Spain. Articles about domestic events (black) and  both  national  and international events (red).}
\vspace*{5mm}
\centering
\includegraphics[scale=0.35]{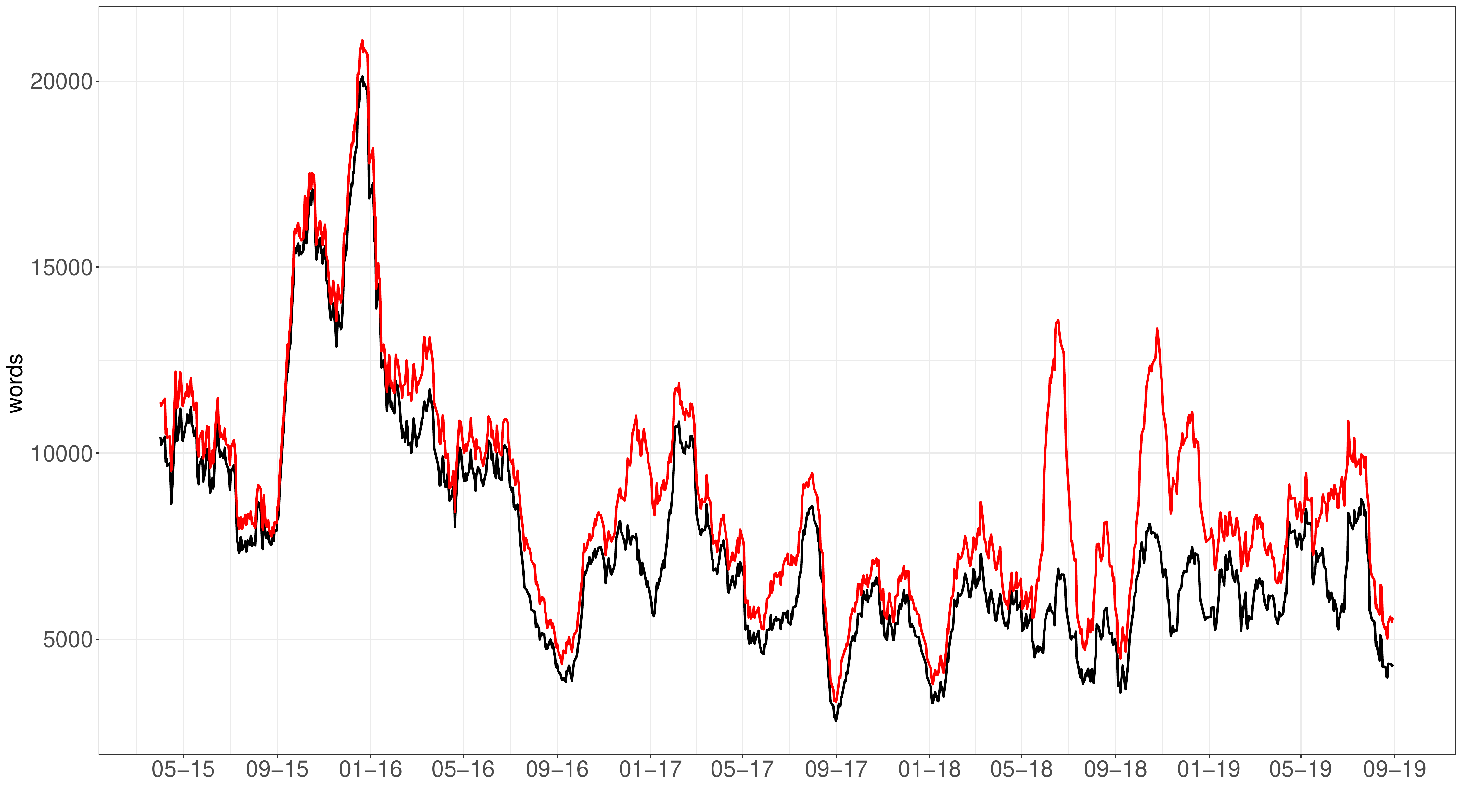}
\end{center}
\label{n_articles_national}
\end{figure}

Figure \ref{n_articles_national} reports the number of articles extracted for the two countries. The red line displays the number of articles mentioning only domestic events, while the black line shows the number of articles having an international focus. Our selection and cleaning procedure leads to a daily average number of articles that ranges between 5 and 25 articles for both countries. The series show a number of peaks at many of the stressful events listed in Table \ref{tab_events}, indicating that the newspapers devote ample discussion to these events and thus validating our selection procedure. The series also show some seasonal patterns, with a considerable reduction in the number of articles extracted during the summer.\footnote{In the period between 18 of July 2018 and 17 August 2018, we observe a sudden drop in the volume of articles collected from the selected Spanish newspapers, by nearly 50 per cent. Further investigation has revealed some problems in the transmission of GDELT data. To mitigate this problem, only for this period of time we have augmented the data set with news from all Spanish newspapers, using the geographic lookup of GDELT sources to determine the nationality of outlets. See https://blog.gdeltproject.org/mapping-the-media-a-geographic-lookup-of-gdelts-sources/ (last access 27 September 2020).} 
It is interesting to note that while for Italy the number of articles on domestic events is very close to the number of articles tackling international events over the entire sample period, for Spain the latter is shifted upwards with respect to the number of articles focusing on domestic events only. Such shift is particularly evident during the second half of the sample period, perhaps indicating that Italian political events have become an international worry, and are the object of many of the extracted articles for Spain.
\newline

\section{Emotions from GDELT economic news} 
\label{s_emotions}

To assess the emotional content of our news, we have adopted the WordNet-Affect emotions classification developed by \cite{Strapparava2004} and \cite{Strapparava2004a}, and mapped automatically within GDELT's news content.
Emotions differ in whether they express a positive or negative overall tone, or valence, as well as on the intensity of the emotional response, also known as \textit{emotional arousal}. 
From the literature in psychology, higher intensity messages affect the comprehension and memorization of the readers since they are often remembered better than neutral ones (see \cite{Megalakaki2019} for a review).
The WordNet-Affect emotion classification scheme is based on the \cite{Ekman1993}’s List of Basic Emotions (i.e. \textit{Anger, Disgust, Fear, Happiness, Sadness, Surprise}) further refined into a set of 32 classes by exploiting the emotional model and categorization proposed by \cite{Elliott1992}. It proposes a lexicon that tries to capture different emotions, or moods, in the text. Under the taxonomy outlined by \cite{Gentzkow2019}, the WordNet-Affect classification is a dictionary based approach that consists of counting the number of terms capturing particular categories of text, a very common approach in the social science literature using text. This emotion classification scheme uses as a starting point the Wordnet, a lexical English database that groups nouns, verbs, adjectives and adverbs into sets of cognitive synonyms (synsets), each expressing a distinct concept (\cite{Miller1995}).\footnote{https://wordnet.princeton.edu (last access 27 September 2020)} Synsets are interlinked by means of conceptual-semantic and lexical relations. Accordingly, \cite{Strapparava2004} have manually produced an initial list of 1,903 terms directly or indirectly referring to mental (e.g. emotional) states (core affective states). Hence, they have exploited the WordNet relations to extend the list of terms expressing such core affective states to obtain a total of 4,787 terms. 
\newline

In this paper we focus on few negative emotions that are often pointed by the financial literature as important in affecting investors' decisions. Specifically, we take \textit{Panic}, representing a state of intense fear or desperation, and \textit{Distress} with the intent to take a state of mild connotation of fear. Under the WordNet-Affect classification, \textit{Panic} is a high-arousal emotion eliciting feelings of strong worry and fright, while \textit{Distress} is associated to words that express worry, concern, uneasiness about some present or future situation, having lower intensity (low arousal). 
The rational for considering these emotions is that we wish to capture the negative affective states of investors, linked to distress and fear when their investments do not perform as expected (\cite{Taffler2018}). We expect intensive negative emotion such as panic to be anticipated by milder conditions of distress and worry, with possibly different impact on the spread.\newline

Table \ref{table_emotions} shows an example of two sentences, highly rated according to our \textit{Distress}, and \textit{Panic} indicators, respectively. In the sentences we report the words belonging to the Loughran and McDonald negative word list highlighted in bold. The table also reports a set of ``secondary'' emotions attached to each primary feelings of Distress and Panic, representing alternative paths to the same emotion. It is interesting to observe that, despite carrying the same overall negative (LM) sentiment, the first two sentences express large differences in the intensity of panic and distress. 
Terms such as ``scare'', ``burn'' or ``fibrillation'' appearing in the second sentence are much more emotionally arousing than milder expressions like ``concern'', ``risky'' present in the first sentence. One possibility to capture such different intensity in negativity of news would be to adopt a graded system that measures how negative/positive a specific term is. However, the calculation of an average, daily, tonality could weep out the effect of specific segments of the polarity. On the contrary, measuring the emotional content of news, can help disentangle the contribution of specific components of the polarity to the total forecasting ability.


\begin{table}[h]
 \begin{center}
 
 \caption{Example of sentences conveying emotions of Distress and Panic}
\label{table_emotions}
 \footnotesize
 \begin{tabular}{lll}
 \\[-1.8ex]\hline 
 \hline \\[-1.8ex] 
 Primary & Secondary \\
 Emotion & Emotion & Example \\
 \\[-1.8ex]\hline 
 \\[-1.8ex]
 \textit{Distress} & Apprehension, Worry, & The signals expressed by the financial markets are beginning \\ 
 & Uneasiness, Concern & to be a source of \textbf{concern} and certainly\\
 & Negative suspense & reflect a fundamental medium-long term risky situation. \\
 \hline
 \textit{Panic} & Fright, & The final bill scares the investors: \textbf{burned} \\ 
 & Shock, Scare, & 12 billion euros in one day \\
 & Terror, Horror, & since the fibrillation began for \\
 & Hysteria & the political situation. \\ 
 \hline
 \end{tabular}
 \end{center}
\footnotesize Notes: The first sentence is taken form https://www.investireoggi.it/obbligazioni/politica-monetaria-e-incertezza-politica-in-italia/, while the second from http://www.ilgiornale.it/news/politica/voto-fa-paura-spread-235-borsa-brucia-tutto-2018-1533515.html.
 \end{table}

For each day in the sample, we calculate the total number of words that carry the negative emotions Distress and Panic appearing in the selected articles published on that day. We standardize these variables by dividing their values by the total number of words in a given day. We then calculate moving averages with a rolling window of 5 open-market days, with the intent to incorporate in our regression model news information referring to the last week. A number of studies also include information on the sentiment for the previous 5 days in their regression (see, for example, \cite{Liu2014}, or \cite{Tetlock_2007} and \cite{Garcia2013}). In particular, for a specific country, we set:

\begin{equation}
 Emotion_{t} = \sum_{s=t-4}^{t} \frac{1}{5} \frac{WC_{emotion,s}}{WC_{s}},
 \label{eq2}
\end{equation}
where $WC_{emotion,s}$ is the words count of the specific emotion for selected articles published in the country at time $s$ according to the Wordnet-Affect lexicon, and $WC_{s}$ is the total words count of all articles published in the day.\footnote{We have done some robustness checks and run several regressions by varying the smoothing parameter in equation (\ref{eq2}) between 3 and 20. Results are similar to those reported in the paper and are available upon request.} 
To facilitate interpretation of regression coefficients, we rescale our indicator to have unit variance.\bigskip
\newline
We next move to our emotion-augmented statistical model for government yield spreads.
\newpage

\section{Methods}
\label{S4}

The main objective of the analysis is to assess how negative emotions impact on the yield spreads during stressed periods. 

Following \cite{Bernal_Gnabo_Guilmin_2016}, we assume that the $q$th percentile of sovereign bond spread expressed in first differences represents a situation of financial distress. Accordingly, let $\Delta Spread_{t + h}^{q}$ be the $q$th percentile of sovereign bond spread for a given country expressed in first differences. We adopt a quantile regression approach and consider the following emotion-augmented quantile regression (\cite{Koenker1978}):
\begin{multline}
 \Delta Spread_{t + 1} = \alpha ^{q} + \delta_{0}^{q} \Delta Spread_{t} + \delta_{1}^{q} X_{t} + \gamma^{q} LM_{t-h} + \beta^{q} Emotion_{t-h} + \epsilon_{t+1}
 \label{eq1}
\end{multline}
where $X_{t}$ includes the variables that are traditionally included in models for government bond spreads to control for various sources of risk, namely credit risk, liquidity risk and risk aversion. $LM_{t}$ is the LM negative indicator expressed as the fraction of the words belonging to the LM negative word list present in the text of the selected articles, and $Emotion_{t}$ is our emotion variable, as defined in (\ref{eq2}). We note that we have included the variable $\Delta Spread_{t}$ amongst the regressors to account for the state of the market.\footnote{We have also tried a specification where we have included the variable $Spread_{t}$ rather than $\Delta Spread_{t}$ amongst the regressors. The main results are similar to those reported in the paper, and hence we have decided not to show them, but are available upon request.}
Further, in our regression we also control for LM with the aim to test whether including our emotion indicators provide any statistically significant incremental information to such general measure of negativity.
In Equation (\ref{eq1}), $h$ is the length of time needed for our news variables to take effect on the dependent variable. In our empirical exercise we try different time horizons, varying from 1 day ($h=1$) up to 1 week lag ($h=5$) for the news to impact on the dynamics of the spread. As for the choice of the quantile level, $q$, following \cite{Bernal_Gnabo_Guilmin_2016}, we focus on the right tail of sovereign bond spread expressed in first differences, and set $q=0.95$. In fact, this measure represents a situation of financial distress where we believe news are most important in anticipating variations in the spread. However, in Section \ref{S5} we also carry a small exploratory analysis to evaluate how the size and significance of estimated regression coefficients attached to the emotion indicators in equation \ref{eq1} vary across quantiles.
\newline

We estimate and evaluate the performance of Model (\ref{eq1}) against the benchmark model containing only financial variables and LM indicator (i.e., where we set $\beta^{q} = 0$) in an in-sample and an out-of-sample exercise. 
In the in-sample analysis, we calculate the $R^{2}$ following the procedure outlined by \cite{Koenker1999}. 
Finally, we compute confidence intervals for the estimated coefficients by using the \cite{Koenker1994} approach based on inversion of a rank test. 
\newline

To evaluate the forecasting performance of Model (\ref{eq1}) against that of the benchmark model, we split the sample into two sub-samples of similar size: we take $T_0=569$ observations for estimation for $h=1$ (567 for $h=5$) and use the remaining observations for testing. Specifically, for each time horizon $h$, and for $t=T_0+1,T_0+2,...,T$ the forecast errors using information up to time $t$ are:
\begin{equation}
 \hat{\epsilon}_{t+1}^{emotion} = \rho_q( \Delta Spread_{t + 1} - \widehat{\Delta Spread}_{t + 1,emotion} )
\end{equation}
\begin{equation}
 \hat{\epsilon}_{t+1}^{benchmark} = \rho_q( \Delta Spread_{t + 1} - \widehat{\Delta Spread}_{t + 1,benchmark} )
\end{equation}
where $\rho_q$ is the so-called \textit{check function}, given by $\rho_q(z) = (q - I(z < 0))z$, $\widehat{\Delta Spread}_{t + 1,emotion}$ is the forecast of the spread using the emotion-augmented quantile regression (\ref{eq1}), and $\widehat{\Delta Spread}_{t + 1,benchmark}$ is the forecast of the spread using the benchmark model.\newline 

Our sample covers a period of high political turmoil, as also evident from Figure \ref{fig1} and Table \ref{tab1}. The large variations in the yield spreads occurring particularly during the second half of the sample period hide important changes in the impact of our regressors on the spread over time, and undermine the validity of standard forecasting tests. As also pointed by \cite{Giacomini2010}, in the presence of structural instability, the relative performance of the two models may itself be time-varying, and thus averaging this evolution over time will result in a loss of information. By selecting the model that performed best on average over a particular historical sample, one may ignore the fact that the competing model produced more accurate forecasts when considering only the recent past. To account for this time variability, in this paper we carry a rolling-windows analysis. Specifically, we re-estimate the unknown parameters in Equation (\ref{eq1}) at each $t=T_0+h,T_0+h+1,...,T$ over a rolling window of $T_0$ days including data indexed $t-h- T_0 +1,...,t-h$, where we set $T_0=569$ observations for estimation for $h=1$ (567 for $h=5$). For each rolling window, we plot the estimated coefficients, confidence intervals and $R^{2}$ against time. For evaluating the forecasting performance of our models, we adopt the Fluctuation test proposed by \cite{Giacomini2010} in order to assess over time, the ability of our news-based emotional indicators to track and anticipate the emotional state of markets over classical predictors during periods of turmoil. The Fluctuation test statistics consists of calculating the \cite{Diebold1995} (DM) test statistics statistic over a rolling out-of sample window of size $m$. When computing this test, we set the ratio between the rolling window length and the out-of-sample length, $\mu=m/(T-T_0)$, equal to $0.30$, as in \cite{Giacomini2010}, and use the HAC estimator for the long-run variance. Finally, in selecting the relevant critical values for the Fluctuation test we consider a two-sided test and assume a nominal size of 5 per cent (see Table I in \cite{Giacomini2010}).
\newpage

\section{Empirical results}
\label{S5}

\subsection{Dynamics of emotions}

\begin{figure}[H]
\begin{minipage}{\textwidth}
    \caption{Distress (green) and panic (red) indicators extracted from GDELT economic news.}
    \label{Appre_Panic}
\vspace*{5mm}
\begin{center}
\subcaption{Italy}
\includegraphics[scale=0.35]{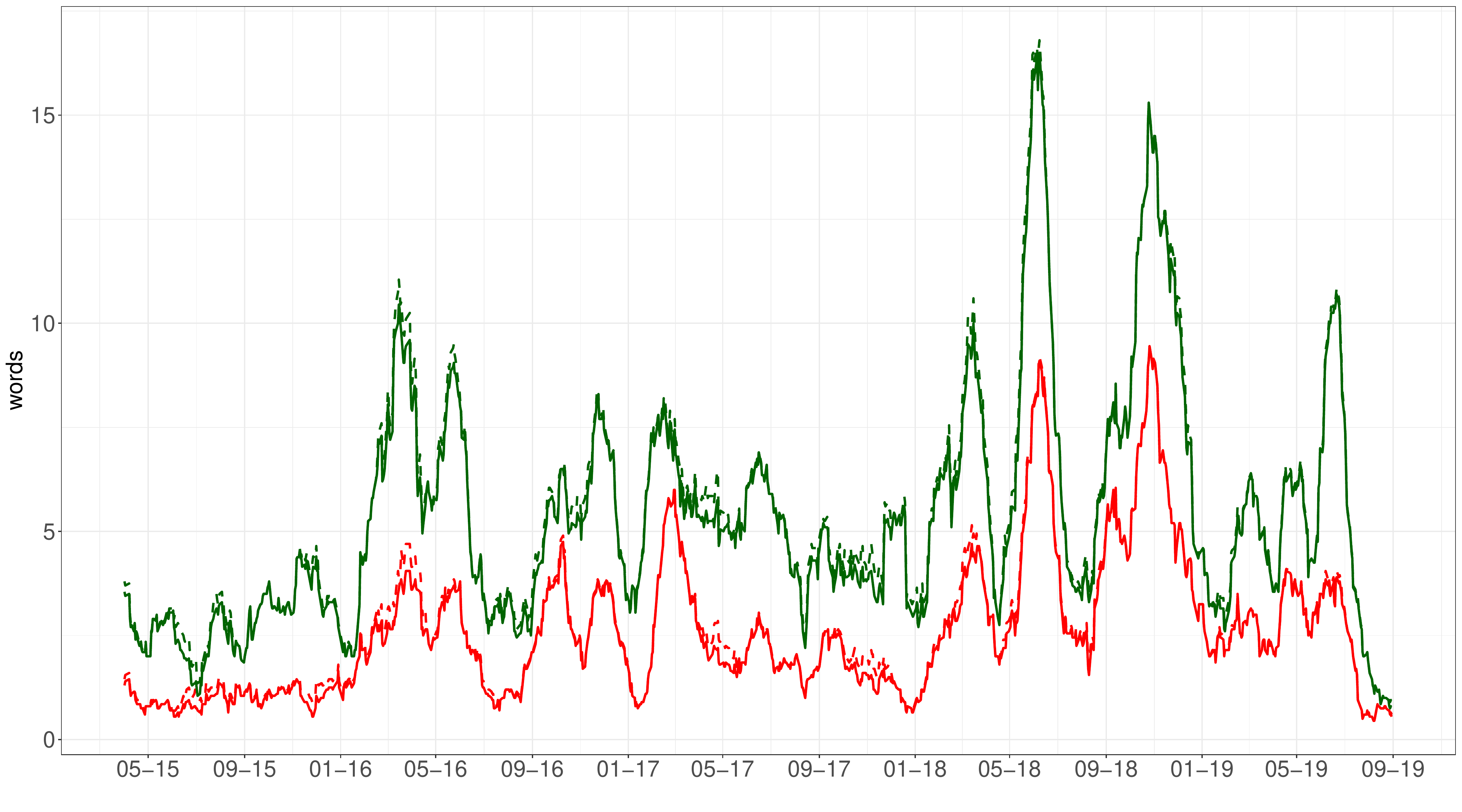}
 \end{center}
\begin{center}
\subcaption{Spain}
\vspace*{5mm}
{\includegraphics[scale=0.35]{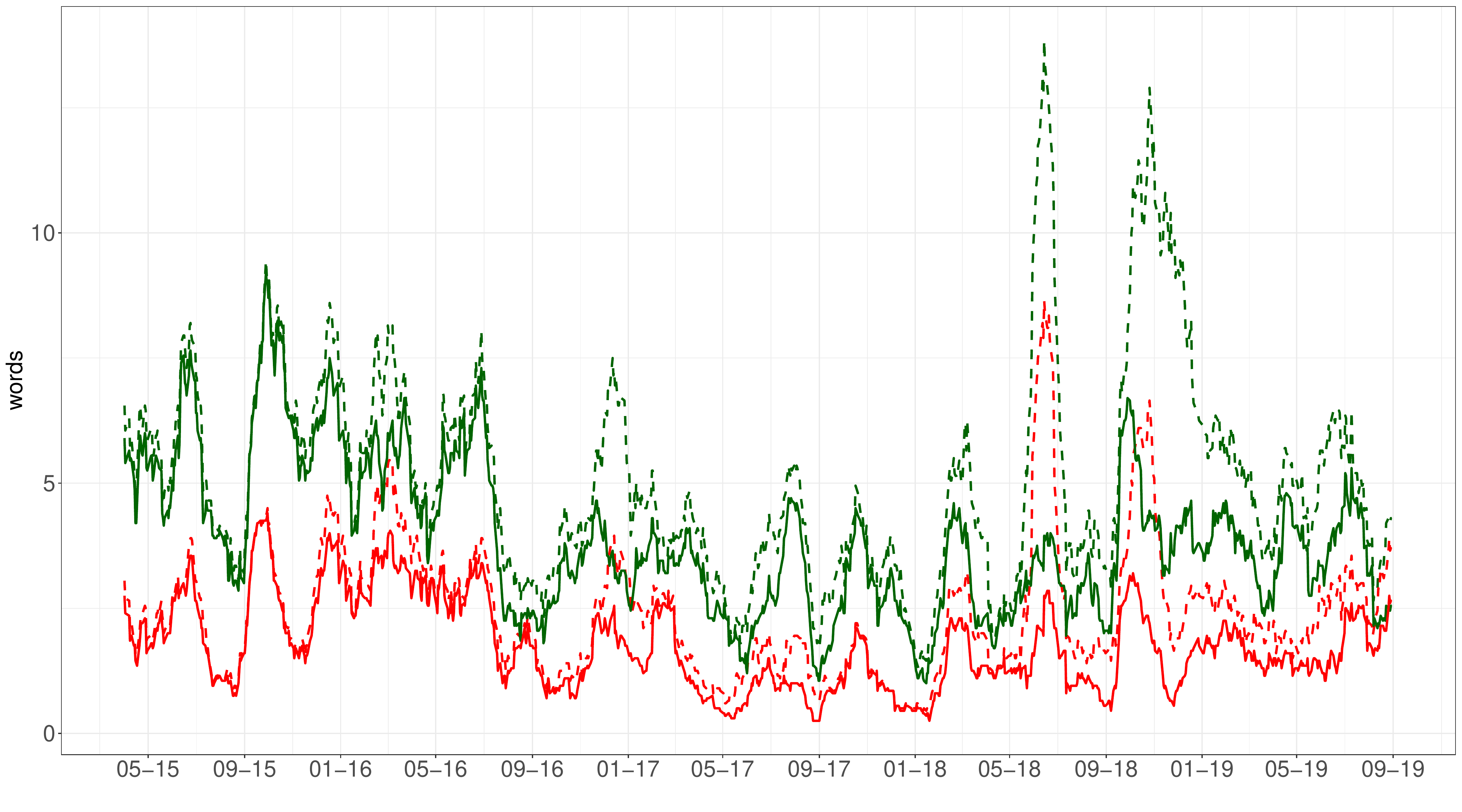}}
\end{center}
\end{minipage}
{\footnotesize Notes: Solid lines refer to news on national events, dashed lines refer to news on national and international events.}
\end{figure}

Figure \ref{Appre_Panic} displays the time series of our emotion variables for Italy (top) and Spain (bottom). The dashed lines in these graphs indicate our emotion indicators calculated on articles that speak about both domestic and international events. Articles extracted carry a daily average of 3 to 18 words conveying Distress and 2 to 8 words carrying Panic, hence, articles contain on average relatively more Distress than Panic. This result could be explained by the fact that milder emotions, are likely to appear more frequently relative to high arousal emotions, such as Panic, that are only elicited in extremely stressing situations. 
The graph shows important time variation in Distress and Panic variables for both Italy and Spain, with the two variables spiking at many of the stressing events listed in Table \ref{tab_events}. Specifically, we observe peaks in the Italian emotions series during the Greece debt crisis in August 2015\footnote{Although in our articles considers predominantly Italian events, the Greece debt crisis emerges as an important issue in this period.}, the Italian constitutional referendum in December 2016 and Italian political elections in March 2018, as well during the period of political turmoil from May to October 2018. Over these months we not only observe spikes but also a rising trend in our emotional variables, indicating an overall increase in negativity. 
As for Spain, the emotion indicators peak during the Spanish general elections (and Brexit Referendum) in June 2016, in occurrence of the October 2017 Catalan referendum, and during the Spanish political crisis in May 2018.\newline

\begin{figure}[H]
\begin{minipage}{\textwidth}
    \caption{LM indicator extracted from GDELT economic news.
    }
    \label{lm_negative}
    \begin{center}
\subcaption{Italy}
{\includegraphics[scale=0.35]{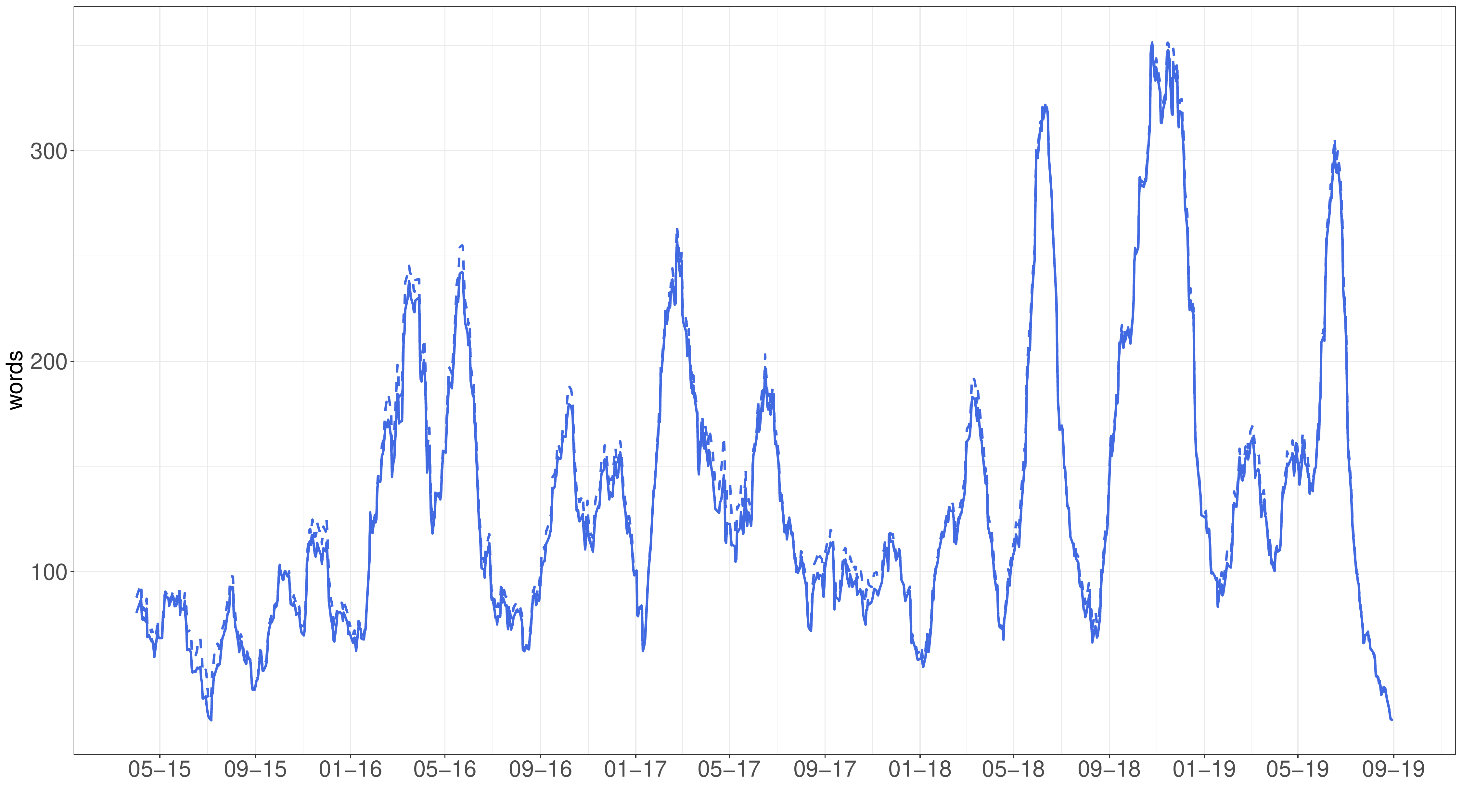}}
\end{center}
\begin{center}
\subcaption{Spain}
{\includegraphics[scale=0.35]{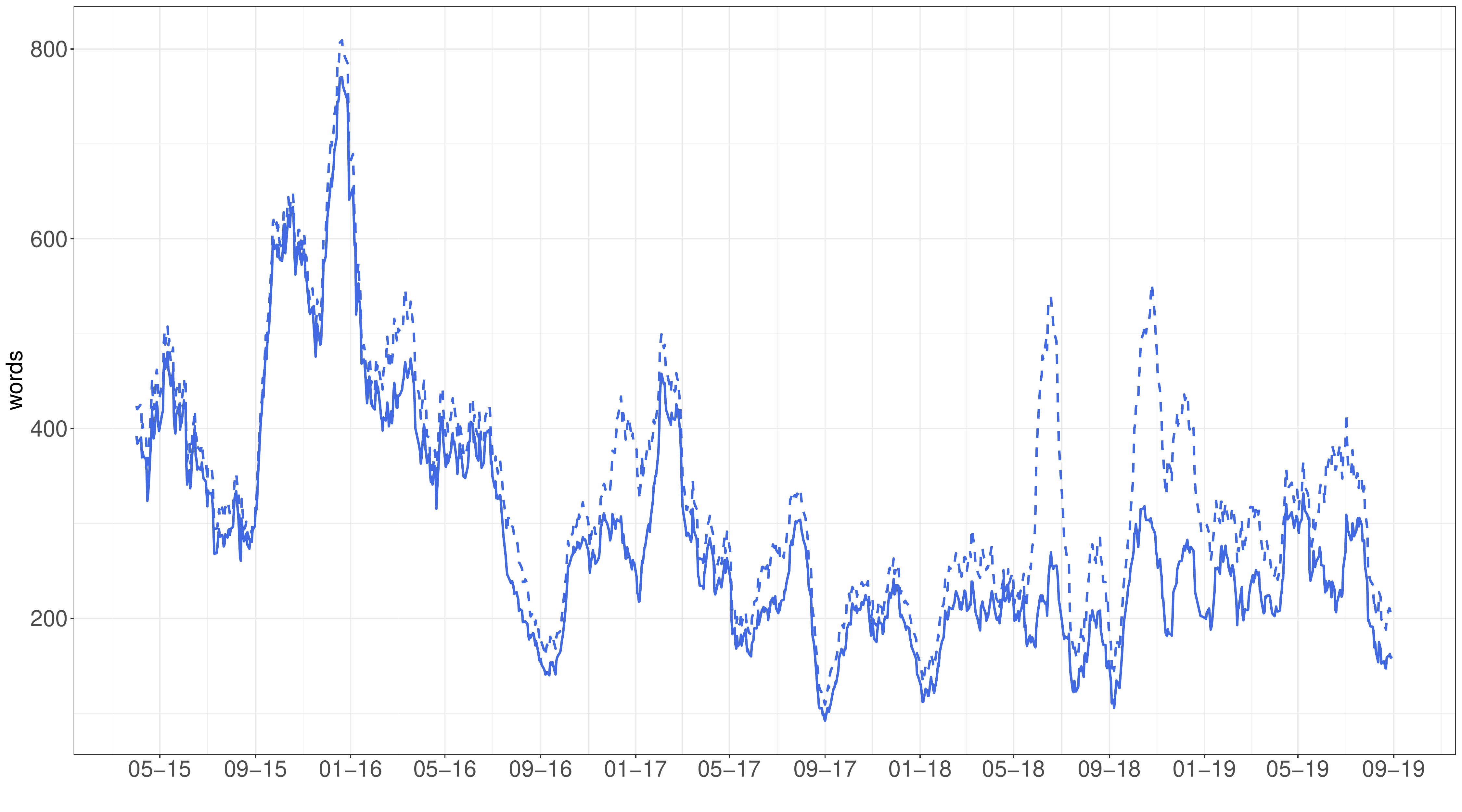}}
\end{center}
\end{minipage}
{\footnotesize Notes: Solid line refers to news on national events, dashed line refers to news on national and international events.}
\end{figure}

Looking at the evolution of the dashed lines in the bottom graph, it is interesting to observe that, during the period of Italian political turmoil and successively, when Italy was discussing deficit spending engagements with the European Union, emotions from Spanish articles speaking about international events considerably diverge from those that focus on domestic facts alone. One explanation for this result is that negative emotions in Spanish national news are mainly triggered by Italian political events.
If we compare the peaks in Distress with those in Panic, we also observe that both series move upwards almost simultaneously, although Distress tends to rise faster than Panic, and Panic seems to revert back to zero more speedily than Distress. This evidence is in line with the financial emotional cycle, where mild worry turns into stronger feelings of fear. 
Figure \ref{lm_negative} shows the temporal evolution of the LM indicator. It is interesting to observe that the behaviour of the LM follows a pattern that is similar to that of our emotional variables, rising in correspondence of the majority of the stressing events previously identified. Another interesting features that emerges from these graphs is that there are on average more negative words in Spanish news stories when international events are taken into account, suggesting attention bias towards what happens abroad.
\newpage

\subsection{Regression analysis}

\begin{figure}[H]
\begin{minipage}{\textwidth}
    \caption{Coefficients and confidence intervals from quantile regression estimation for Italy and Spain, when varying the quantile level $q$ between 0.05 and 0.95.}
    \label{explore_quant}
    \begin{center}
\subcaption{Italy}
\includegraphics[scale=0.35]{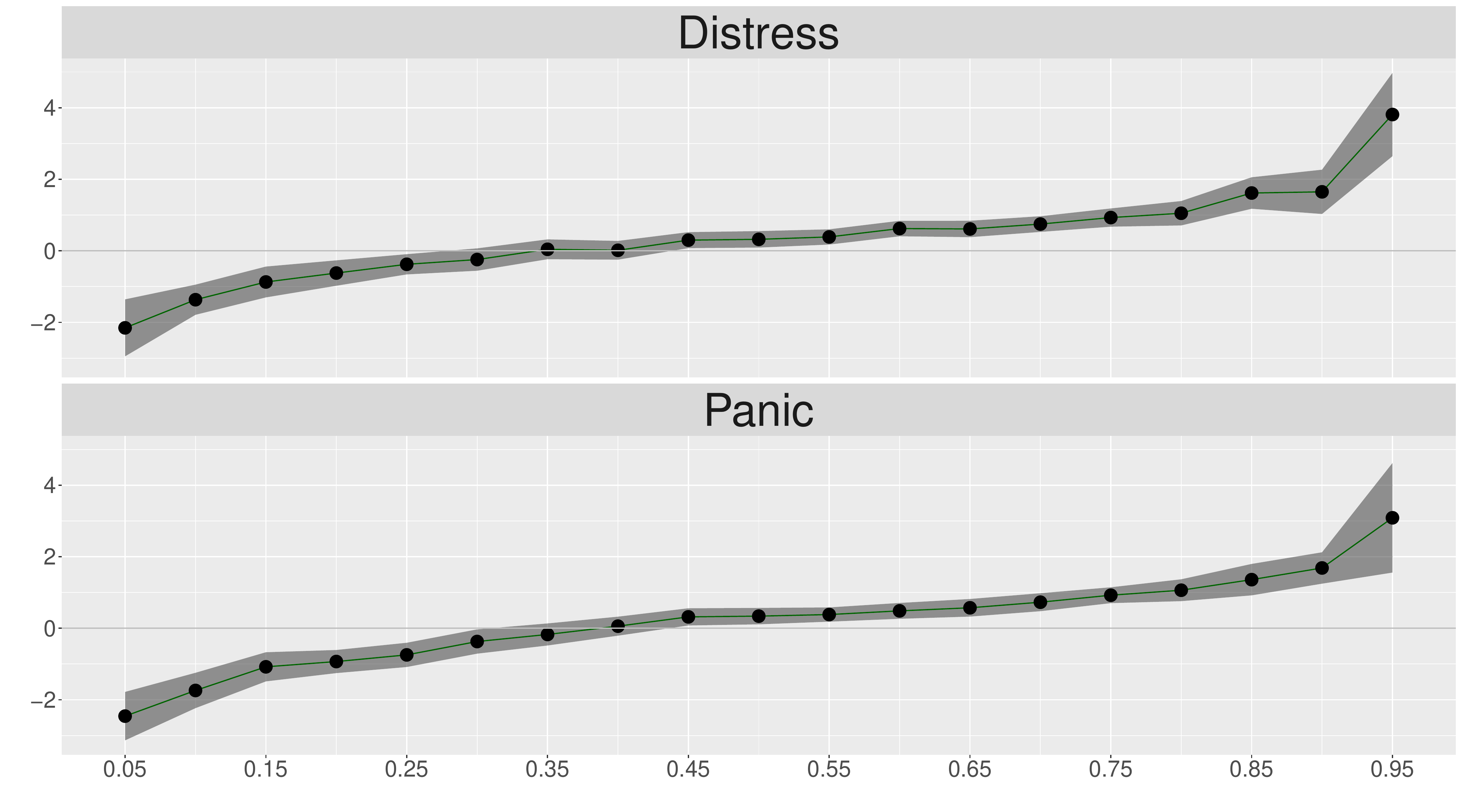}
\end{center}
\begin{center}
\subcaption{Spain}
\includegraphics[scale=0.35]{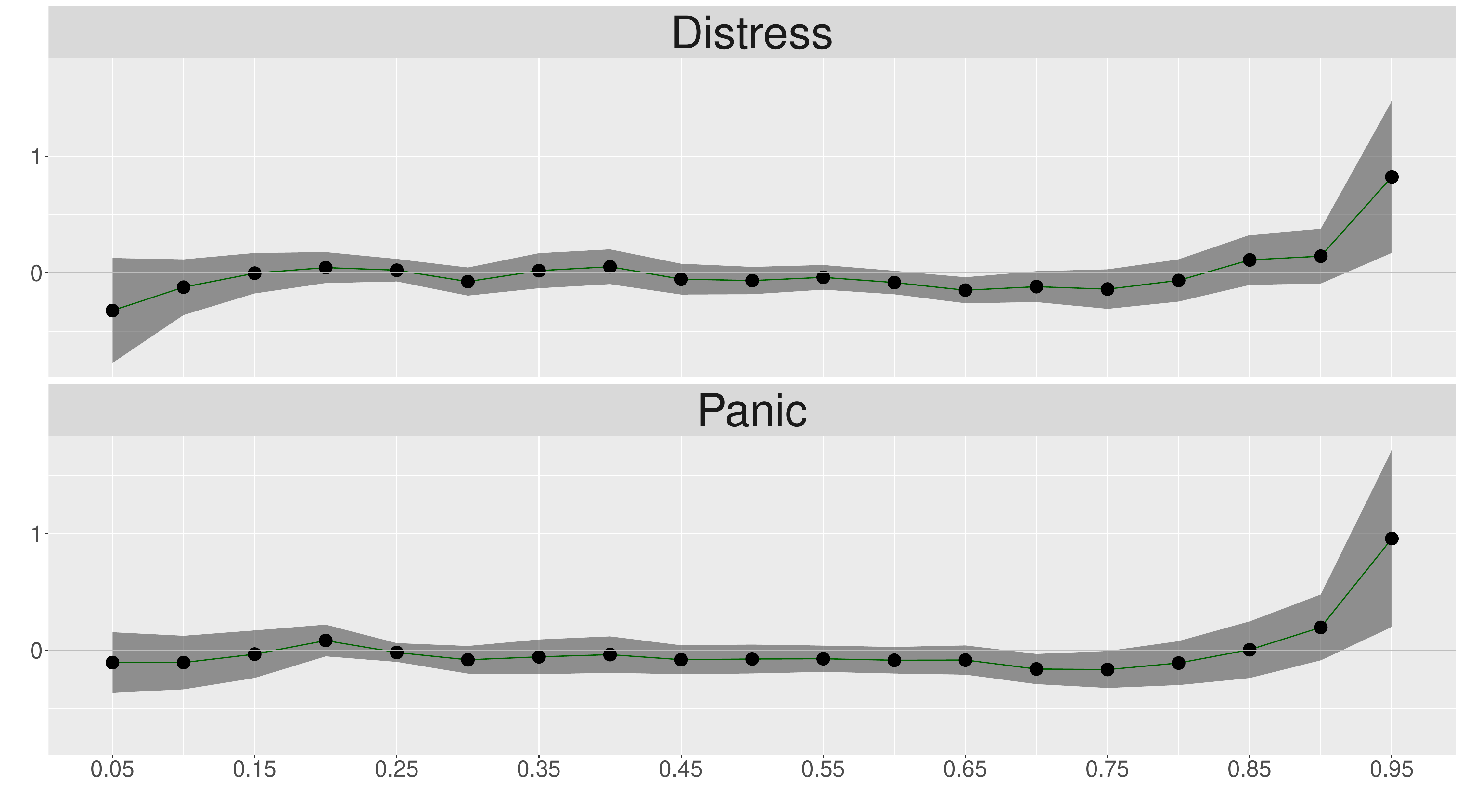}
\end{center}
\end{minipage}
{\footnotesize Notes: In the graphs we report estimated coefficients and associated confidence intervals for $\hat \beta^{q}$ on from Equation \ref{eq1}: $\Delta Spread_{t + 1} = \alpha ^{q} + \delta_{0}^{q} \Delta Spread_{t}  + \delta_{1}^{q} X_{t} + \gamma^{q} LM_{t-h} + \beta^{q} Emotion_{t-h} + \epsilon_{t+1}$ with $q=0.05,0.1,0.15,...,0.95$, for $h=0$.}
\end{figure}

Figure \ref{explore_quant} displays the size and significance of the estimated coefficients attached to our news indicators Panic and Distress in (\ref{eq2}) when varying the parameter $q$ between 0.05 and 0.95 at intervals of 0.05. It is interesting to observe that the news indicators are significantly different from zero for $q$ that lies either near zero or near 1, while they are not significant for quantiles around 0.5. This result seems to indicate that news are most important in anticipating variations in the spreads particularly during period of high uncertainty and distress, and support the choice of $q=0.95$ in the estimation of equation (\ref{eq2}). Accordingly, the rest of the analysis focuses in the case $q=0.95$. We now turn to the rolling-window regression analysis.\newline 

\begin{figure}[H]
    \caption{Rolling coefficients and confidence intervals from quantile regression estimation for Italy and Spain (focus on Domestic events) with $h=0$ (top) $h=1$ (center) $h=5$ (bottom).}
    \label{coeff_h_0_1_5_IT_SP_Domestic}
    \begin{center}
    \subcaption{Italy}
    \includegraphics[scale=0.35]{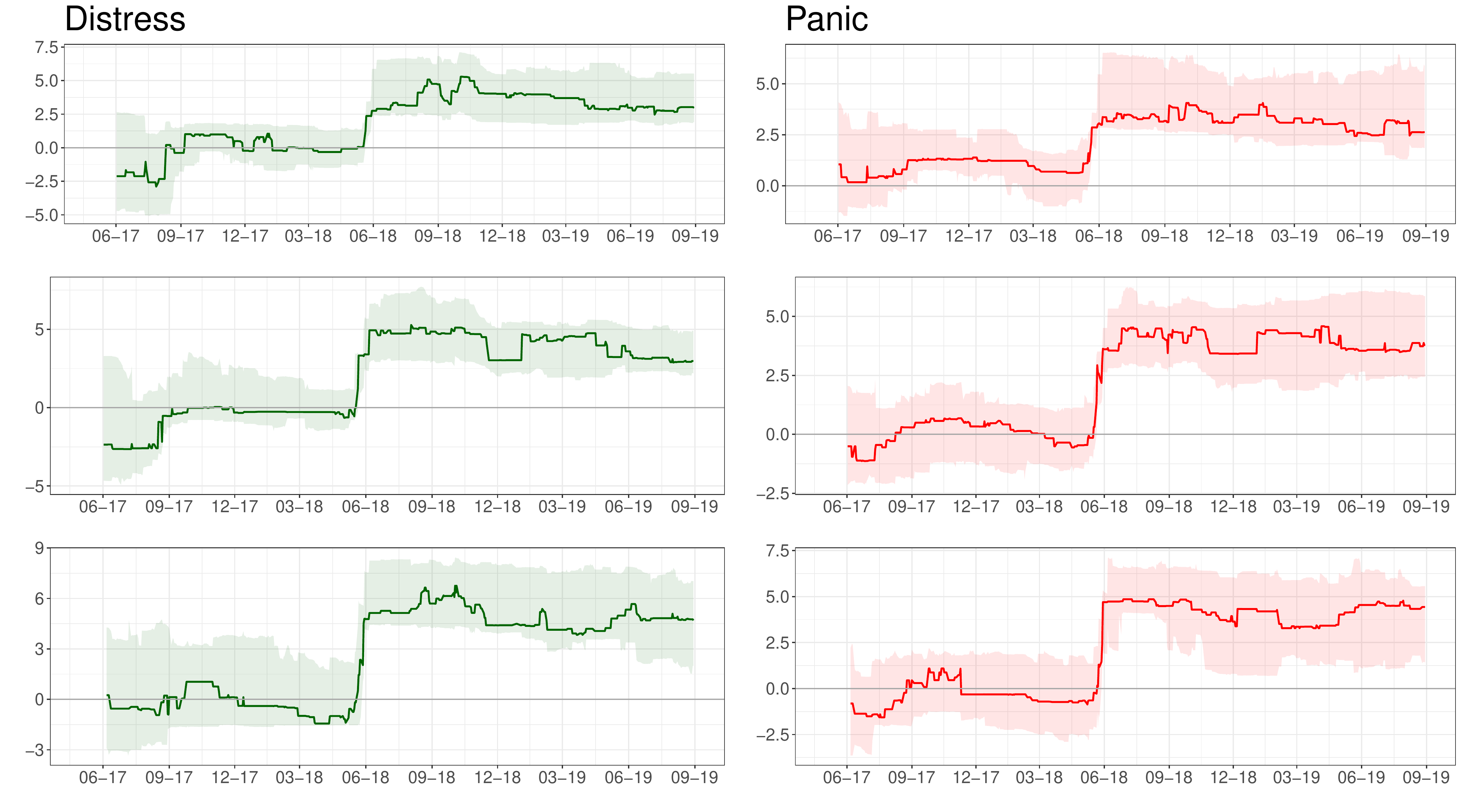}
    \end{center}
    \begin{center}
    \subcaption{Spain}
     \includegraphics[scale=0.35]{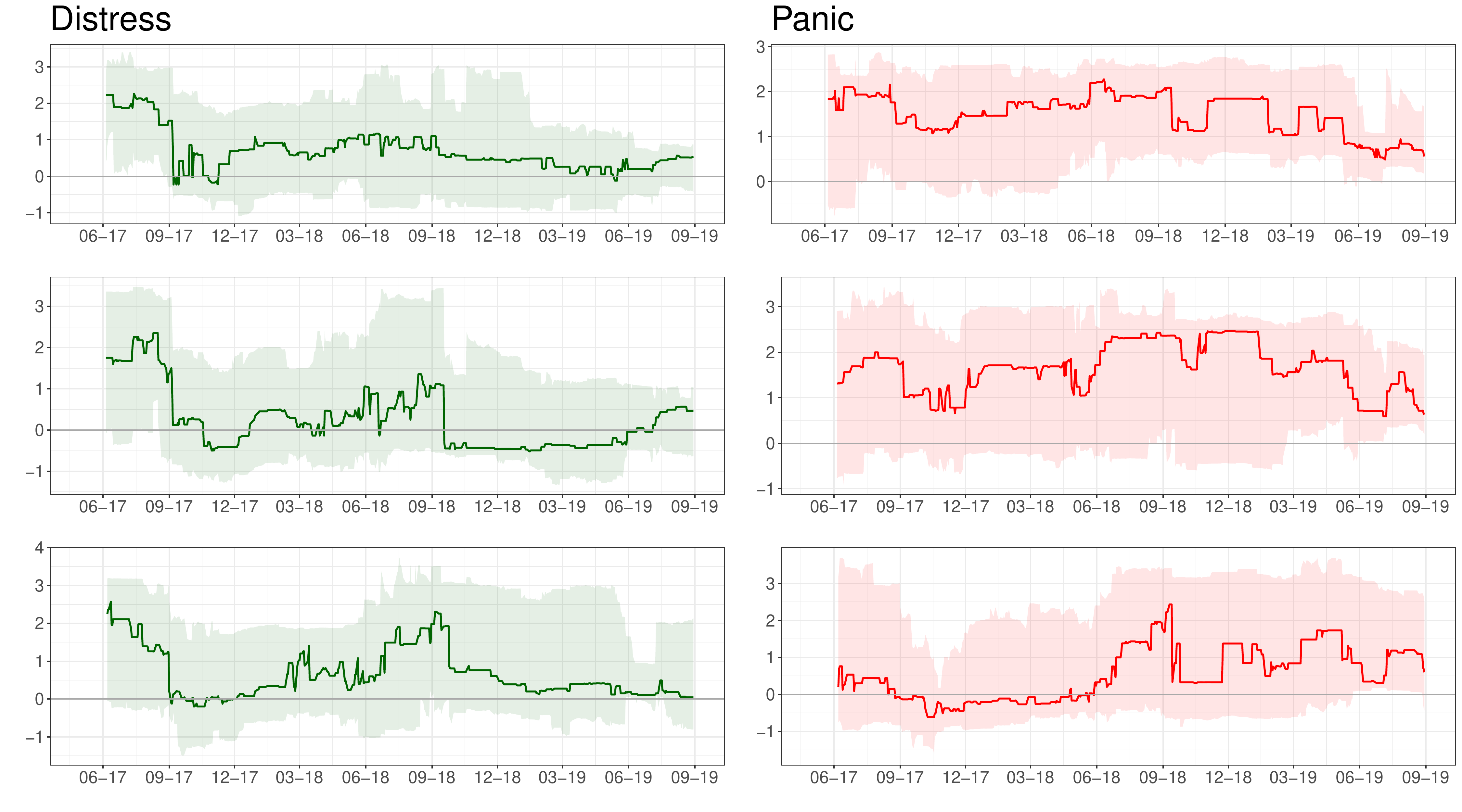}
     \end{center}
{\footnotesize Notes: In the graphs we report regression coefficients $\hat \beta^{q}$ and associated confidence intervals evaluated over rolling windows from Equation (\ref{eq1}): $\Delta Spread_{t + 1} = \alpha ^{q} + \delta_{0}^{q} \Delta Spread_{t}  + \delta_{1}^{q} X_{t} + \gamma^{q} LM_{t-h} + \beta^{q} Emotion_{t-h} + \epsilon_{t+1}$ with $q=0.95$.}
\end{figure}

\begin{figure}[H]
    \caption{Rolling coefficients and confidence intervals from quantile regression estimation for Italy and Spain (Domestic and International events) with $h=0$ (top) $h=1$ (center) $h=5$ (bottom).}
    \label{coeff_h_0_1_5_IT_SP_International}
   \begin{center}
    \subcaption{Italy}
    \includegraphics[scale=0.35]{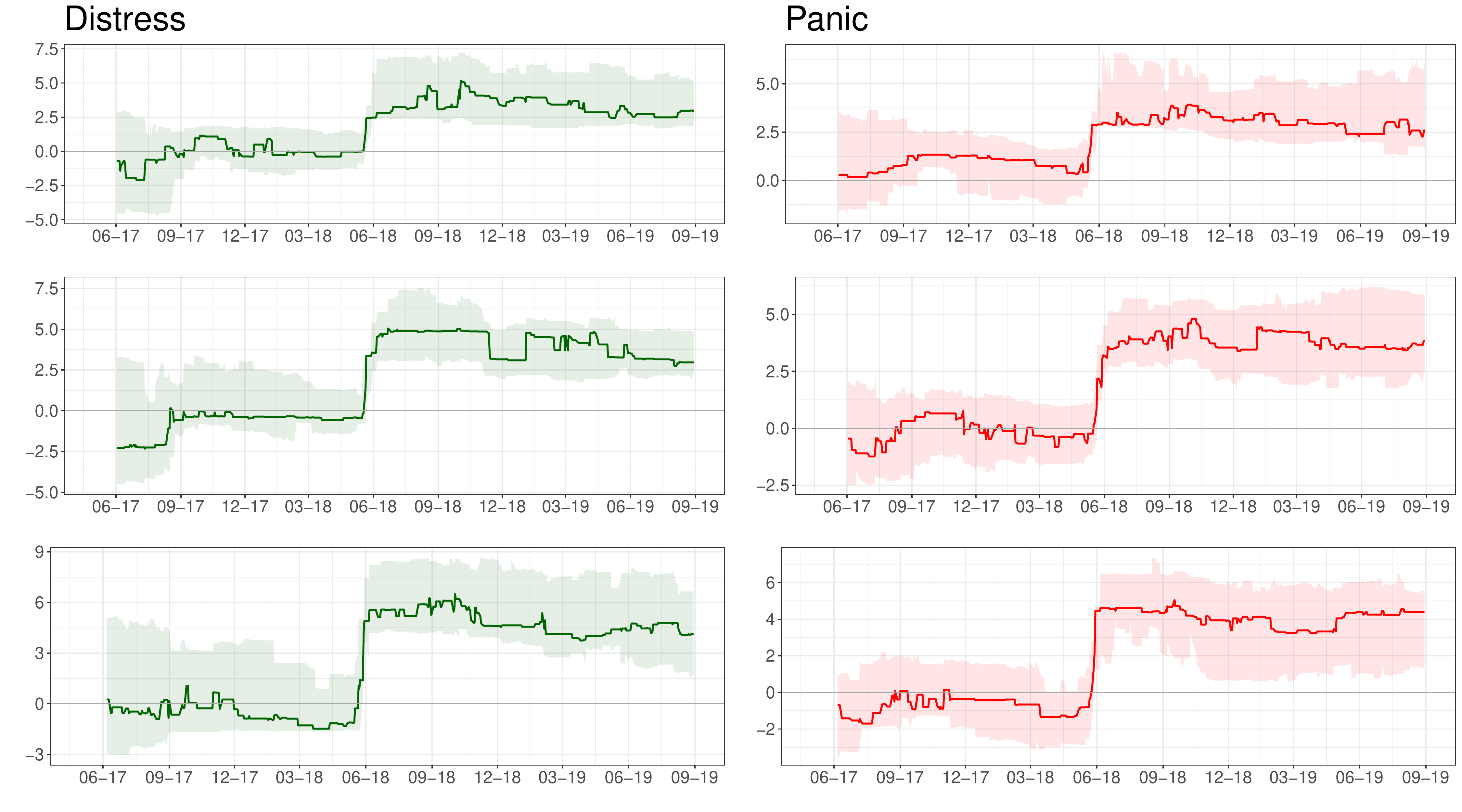}
    \end{center}
 \begin{center}
    \subcaption{Spain}
   \includegraphics[scale=0.35]{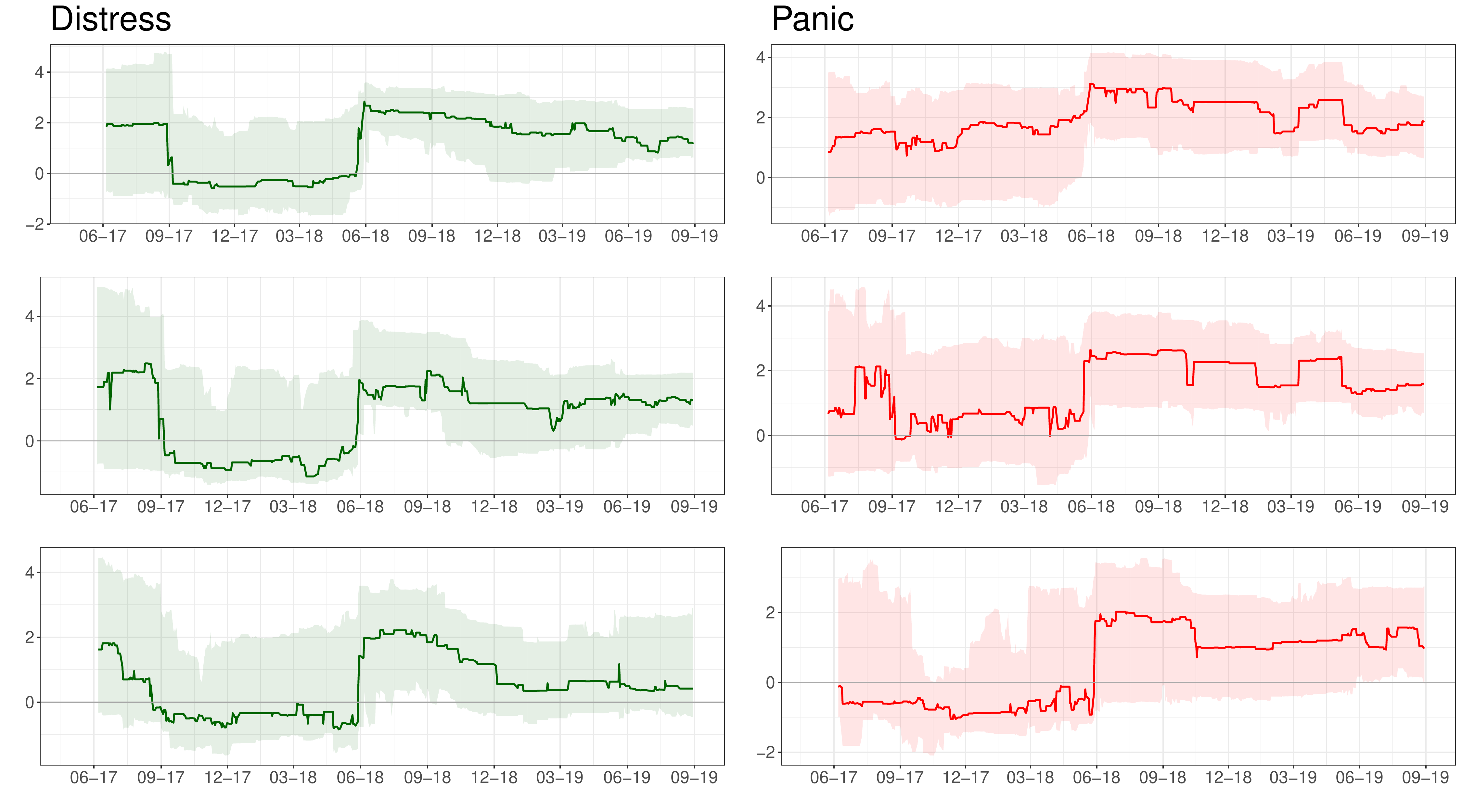}
    \end{center}
 {\footnotesize Notes: In the graphs we report regression coefficients $\hat \beta^{q}$ and associated confidence intervals evaluated over rolling windows from Equation (\ref{eq1}): $\Delta Spread_{t + 1} = \alpha ^{q} + \delta_{0}^{q} \Delta Spread_{t}  + \delta_{1}^{q} X_{t} + \gamma^{q} LM_{t-h} + \beta^{q} Emotion_{t-h} + \epsilon_{t+1}$ with $q=0.95$.}

\end{figure}

Figure \ref{coeff_h_0_1_5_IT_SP_Domestic} visualises the evolution over the rolling windows of the estimated coefficients from Equation (\ref{eq1}) and associated confidence intervals. It shows results for Italy and Spain when setting $h=0,1,5$ and the news focus on domestic events, while Figure \ref{coeff_h_0_1_5_IT_SP_International} show the same graphs when the focus is on domestic and international events.\footnote{For ease of exposition, results for the case $h=2,3,4$ are not reported but are available upon request.}
The rolling-window estimation uncovers important time variations both in the size and significance of coefficients. For Italy, we observe that starting from the beginning of the political crisis in May 2018, all estimated coefficients turn strongly significant for all values of $h$ until the end of the sample. Over this period of time, the estimated coefficients attached to our emotion variables are always positive. Specifically, one standard deviation shock to the Distress variable produces a change in the spread that range between 2.5 and 6 basis points, while such change ranges between 2.5 and 5 basis points for a shock of the same amplitude on the Panic indicator.
For all indicators, the coefficients tend to rise when using larger values of $h$, although the associated confidence bands also tend to widen. 
Similar results are obtained when extending the focus to international events, indicating that the predictive performance of the Italian news is mainly driven by concern and scare triggered by domestic events. This result is somewhat expected, given that for Italy, as evident from Figure \ref{Appre_Panic}, our news indicators are almost identical when focusing on domestic or international events.\newline

Moving to Spain, Figure \ref{coeff_h_0_1_5_IT_SP_Domestic} shows that when focusing on domestic news, only the coefficients attached to Panic, for $h=0$ (and to some extend for $h=1$) are statistically significant. 
However, it is interesting to observe from Figure \ref{coeff_h_0_1_5_IT_SP_International} that when expanding the focus to international events, Panic and Distress turn strongly significant in May 2018. This result seems to support the view that for Spain, the power of news in predicting variations in government spread is mostly explained by emotions triggered by international rather than domestic events. The effects of one standard deviation shock to our emotion variables produces a change on the Spanish spread that range between 1 and 2 basis points with values that can reach 3 basis points for Panic at short time horizons. While the statistical significance of parameters attached to Panic is maintained till the end of the sample, for Distress it is mainly concentrated in the second part of 2018. We also notice a deterioration of the forecasting performance at longer horizons. While for $h=0,1$ the parameters attached to Panic are strongly significant over a long interval of time, it turns insignificant for $h=5$.  We also observe that the values of the estimated coefficients are overall smaller relative to the Italian case.\newline 

\begin{figure}[H]
\caption{Rolling $R^2$ for Italy (Domestic events) with $h=0$ (top) $h=1$ (center) $h=5$ (bottom).}
\label{R2_Italy_domestic}
\begin{center}
\includegraphics[scale=0.22]{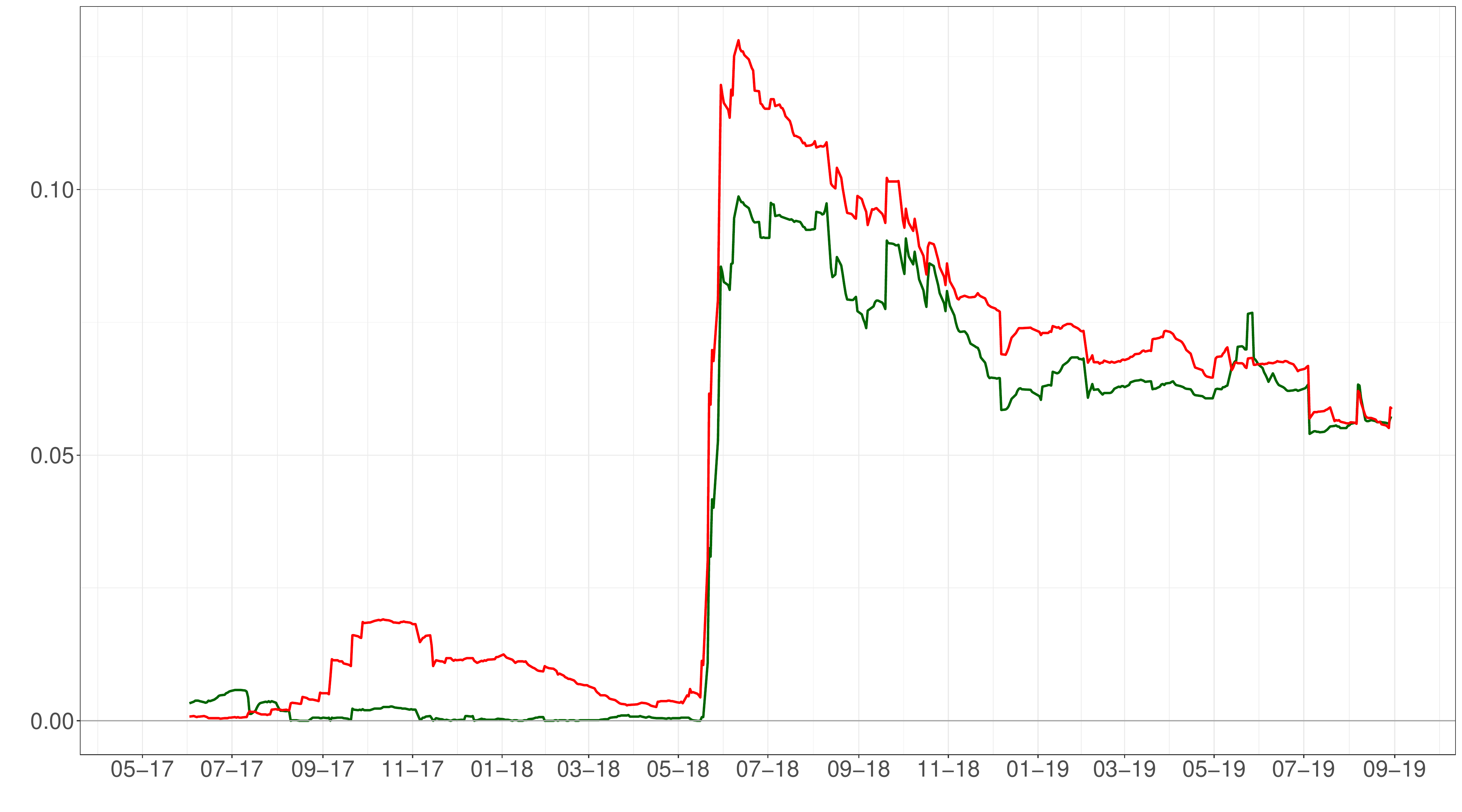}
\end{center}
\begin{center}
\includegraphics[scale=0.22]{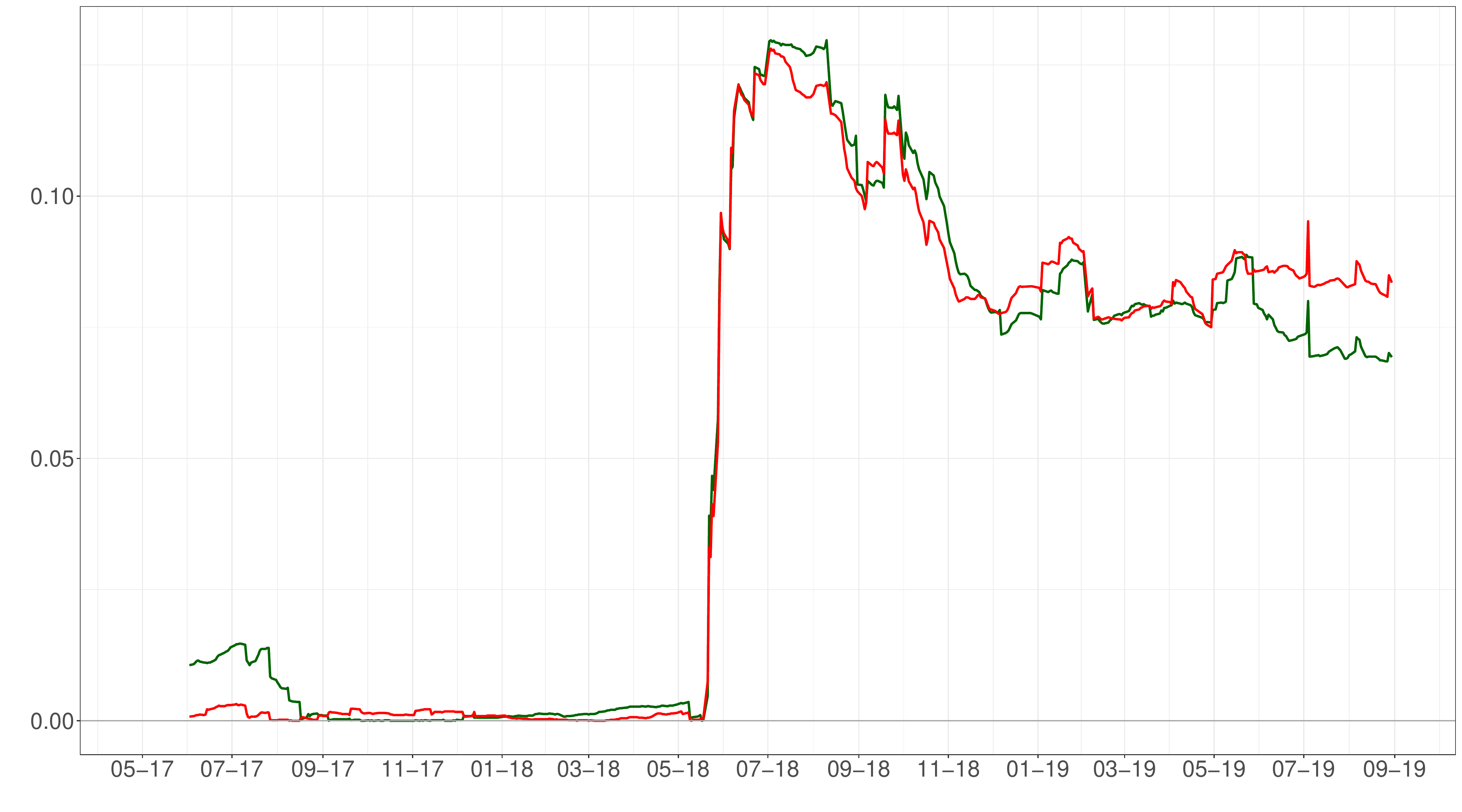}
\end{center}
\begin{center}
\includegraphics[scale=0.22]{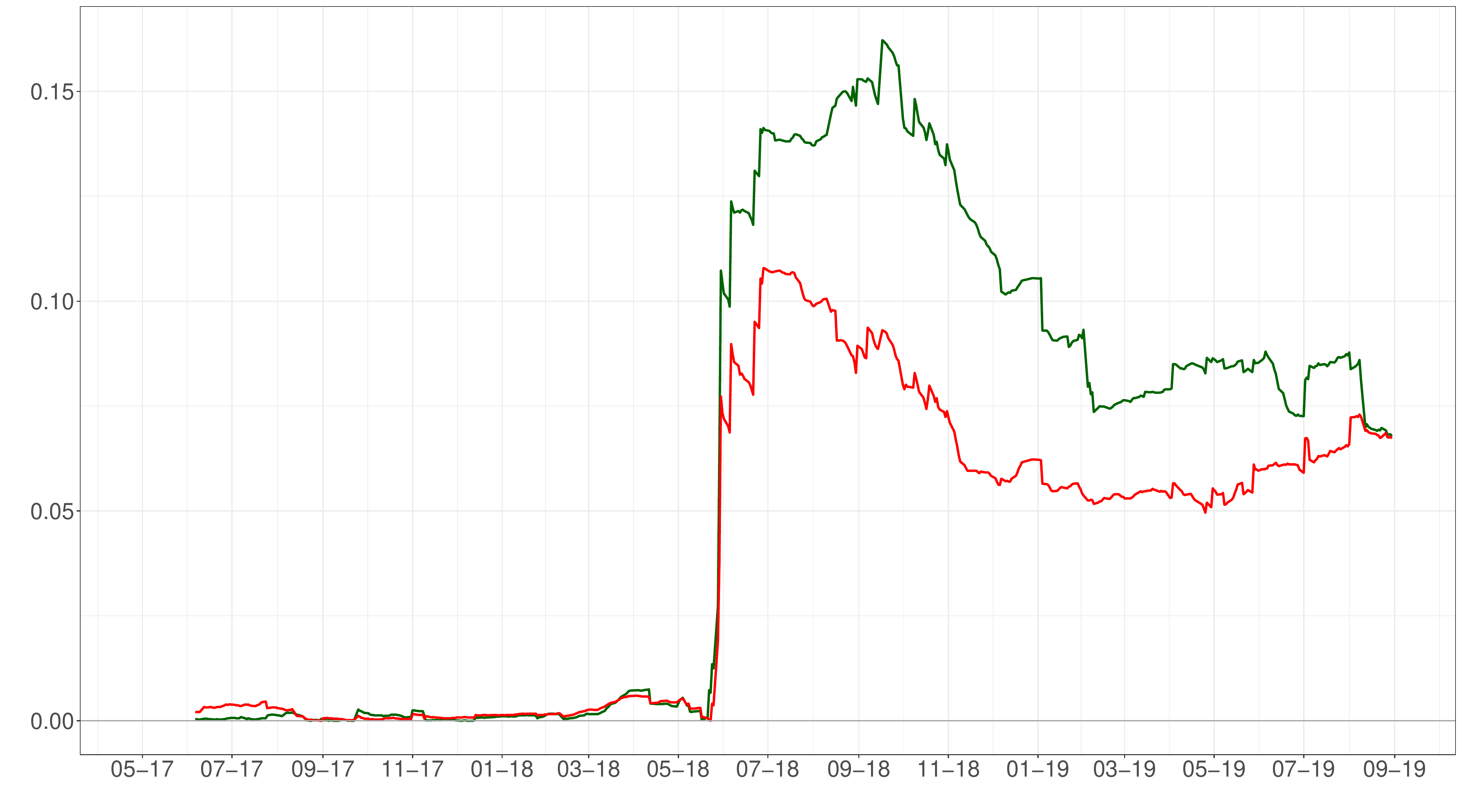}
\end{center}
{\footnotesize Notes: The rolling $R^2$ displayed are calculated as a difference of the $R^2$ for the regression $\Delta Spread_{t + 1} = \alpha ^{q} + \delta_{0}^{q} \Delta Spread_{t}  + \delta_{1}^{q} X_{t} + \gamma^{q} LM_{t-h} + \beta^{q} Emotion_{t-h} + \epsilon_{t+1}$ from the $R^2$ of the benchmark model $\Delta Spread_{t + 1} = \alpha ^{q} + \delta_{0}^{q} \Delta Spread_{t}  + \delta_{1}^{q} X_{t} + \gamma^{q} LM_{t-h} + \epsilon_{t+1}$ where $\beta^{q}$ is set to zero and with $q=0.95$. The red line is for $Emotion_{t}=Panic_{t}$, the green line is  $Emotion_{t}=Distress_{t}$.}

\end{figure}

\begin{figure}[H]
\centering
\begin{minipage}{\textwidth}
    \caption{Rolling $R^2$ for Spain (Domestic and international events) with $h=0$ (top) $h=1$ (center) $h=5$ (bottom).}
    \label{R2_Spain_domestic_international}
\begin{center}
\includegraphics[scale=0.22]{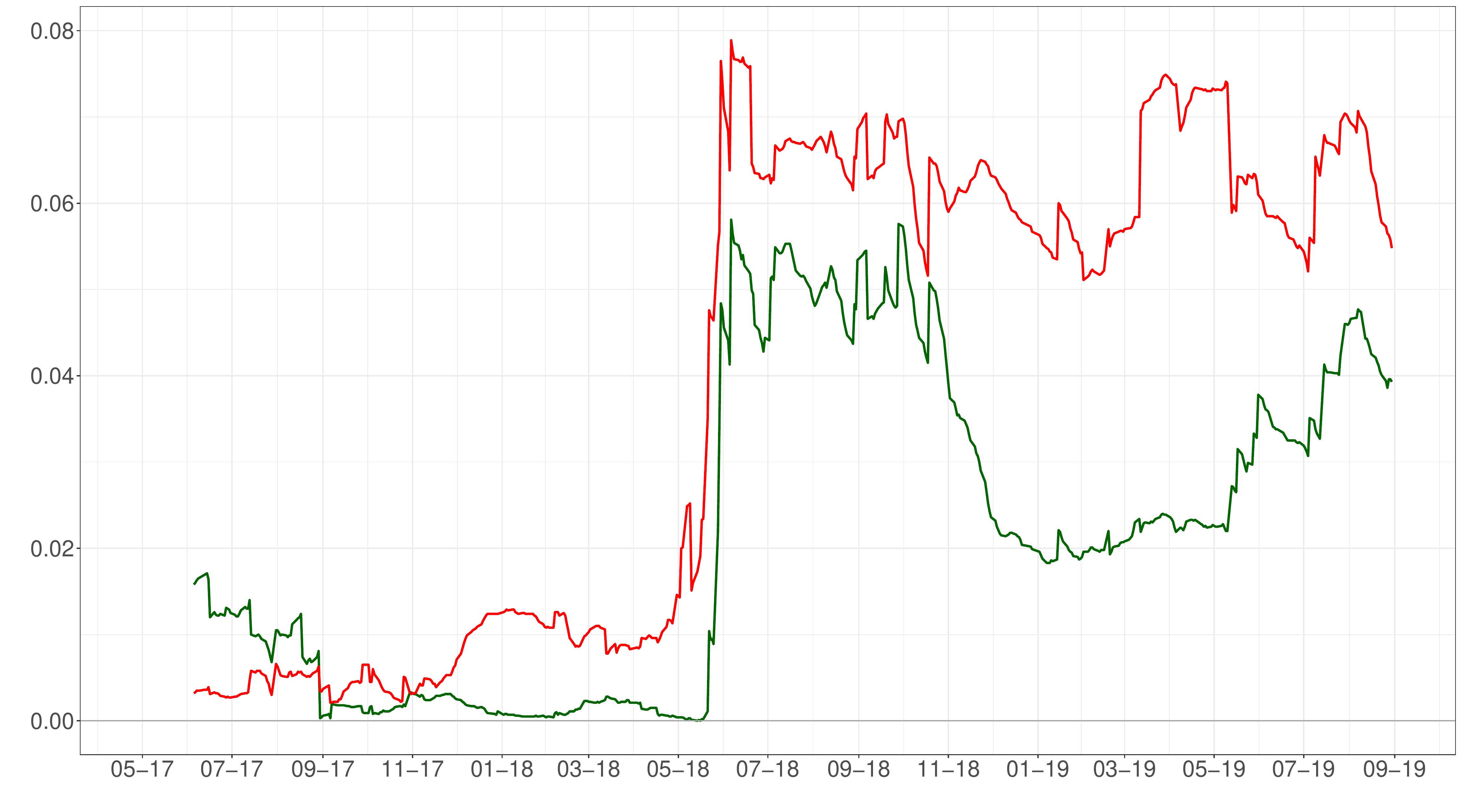}
\end{center}
\begin{center}
\includegraphics[scale=0.22]{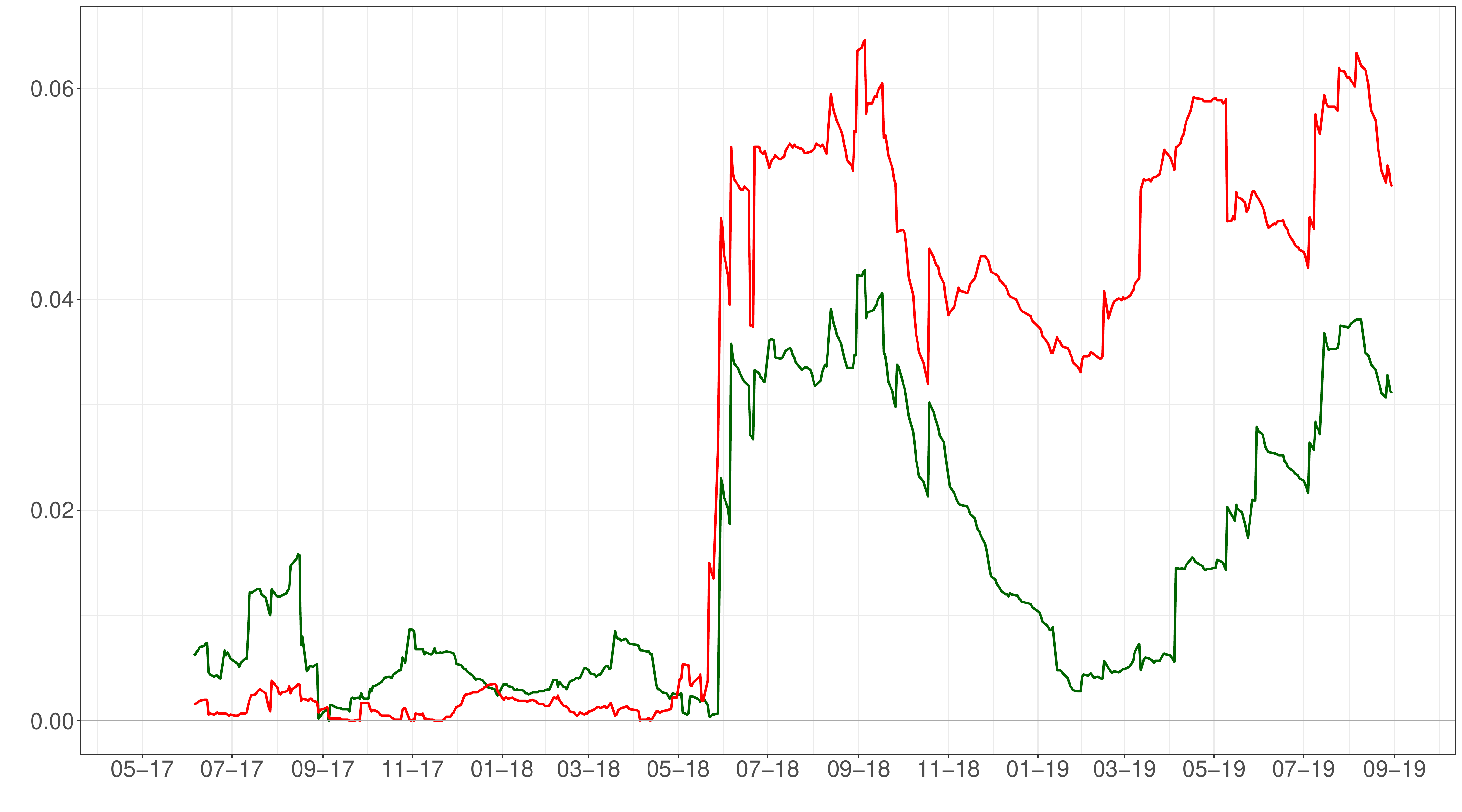}
\end{center}
\begin{center}
\includegraphics[scale=0.22]{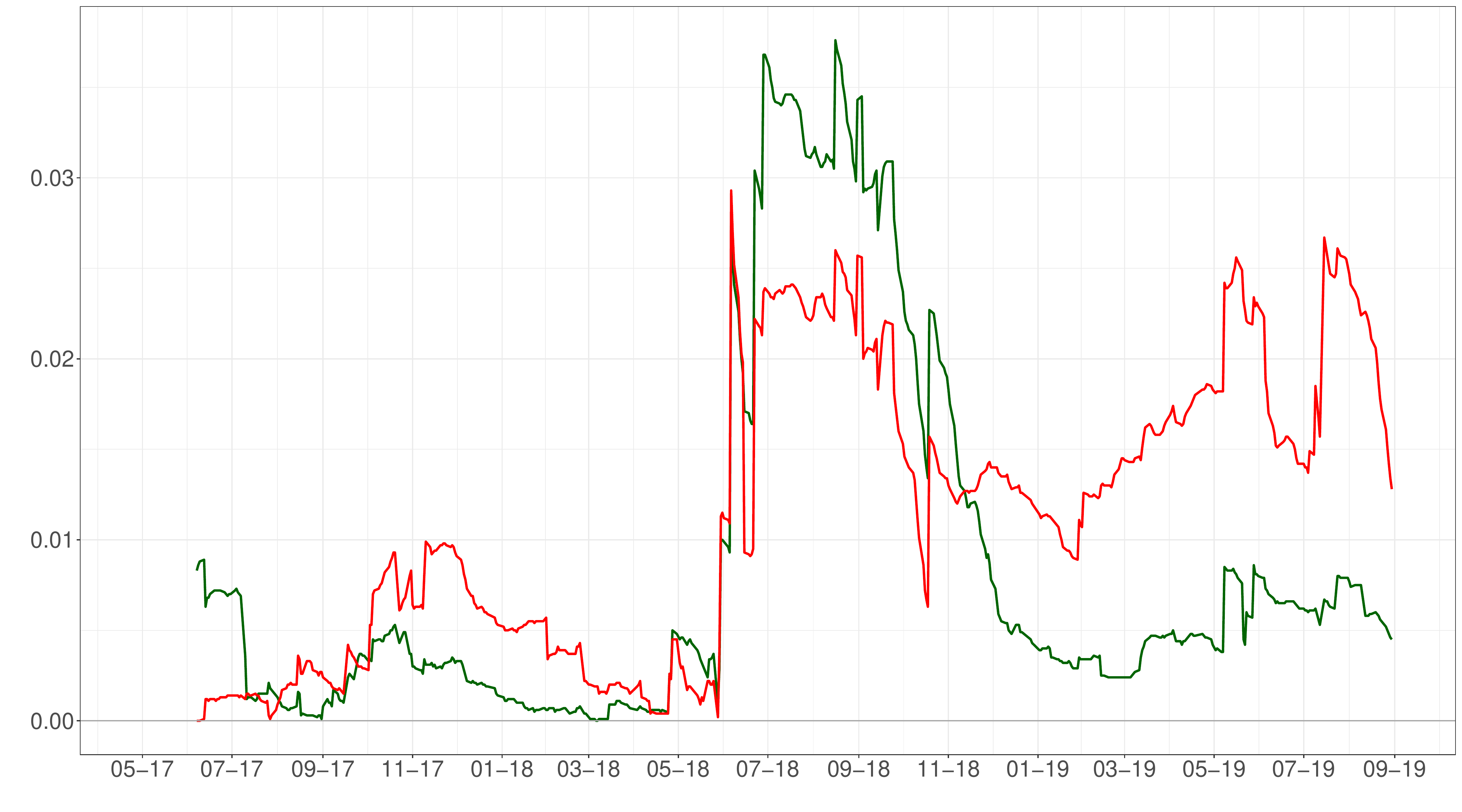}
\end{center}
{\footnotesize Notes: The rolling $R^2$ displayed are calculated as a difference of the $R^2$ for the regression $\Delta Spread_{t + 1} = \alpha ^{q} + \delta_{0}^{q} \Delta Spread_{t}  + \delta_{1}^{q} X_{t} + \gamma^{q} LM_{t-h} + \beta^{q} Emotion_{t-h} + \epsilon_{t+1}$ from the $R^2$ of the benchmark model $\Delta Spread_{t + 1} = \alpha ^{q} + \delta_{0}^{q} \Delta Spread_{t}  + \delta_{1}^{q} X_{t} + \gamma^{q} LM_{t-h} + \epsilon_{t+1}$ where $\beta^{q}$ is set to zero and with $q=0.95$. The red line is for $Emotion_{t}=Panic_{t}$, the green line is  $Emotion_{t}=Distress_{t}$.}
\end{minipage}
\end{figure}

Figures \ref{R2_Italy_domestic} and \ref{R2_Spain_domestic_international} plots the difference in adjusted $R^2$ between the models with our news indicators and the benchmark over the rolling windows. For ease of exposition, in the rest of the paper we only report the case of domestic news for Italy and the case of domestic and international news for Spain. In correspondence of the Italian political crisis in May 2018, the in-sample performance of the model improves. In particular, while Distress seems to boost the in-sample performance at longer forecasting horizons (i.e., $h=5$), Panic is more important in explaining contemporaneous reactions of Italian spread (i.e., for $h=0$). Hence, according to our results, milder emotions seem to capture the stressed state of the market with predictive power at longer time horizons, as opposed to stronger emotions that have important instantaneous effects. This result supports the view that milder emotions trigger stronger ones that are then associated to instantaneous variations in the spread. For Spain, Panic is the emotion that mostly contributes to the in-sample fitting performance for $h=0,1$, while Distress is more important at $h=5$, although the rolling $R^2$ quickly reduces after the peak in the May 2018. Clearly, the Spanish spread is driven by strong emotions triggered by international events.\bigskip \newline

\begin{figure}[H]
\begin{minipage}{\textwidth}
    \caption{Fluctuation test for Italy (Domestic events) with $h=1$ (top) $h=5$ (bottom).}
    \label{fluctuationitaly}
    \begin{center}
\includegraphics[scale=0.25]{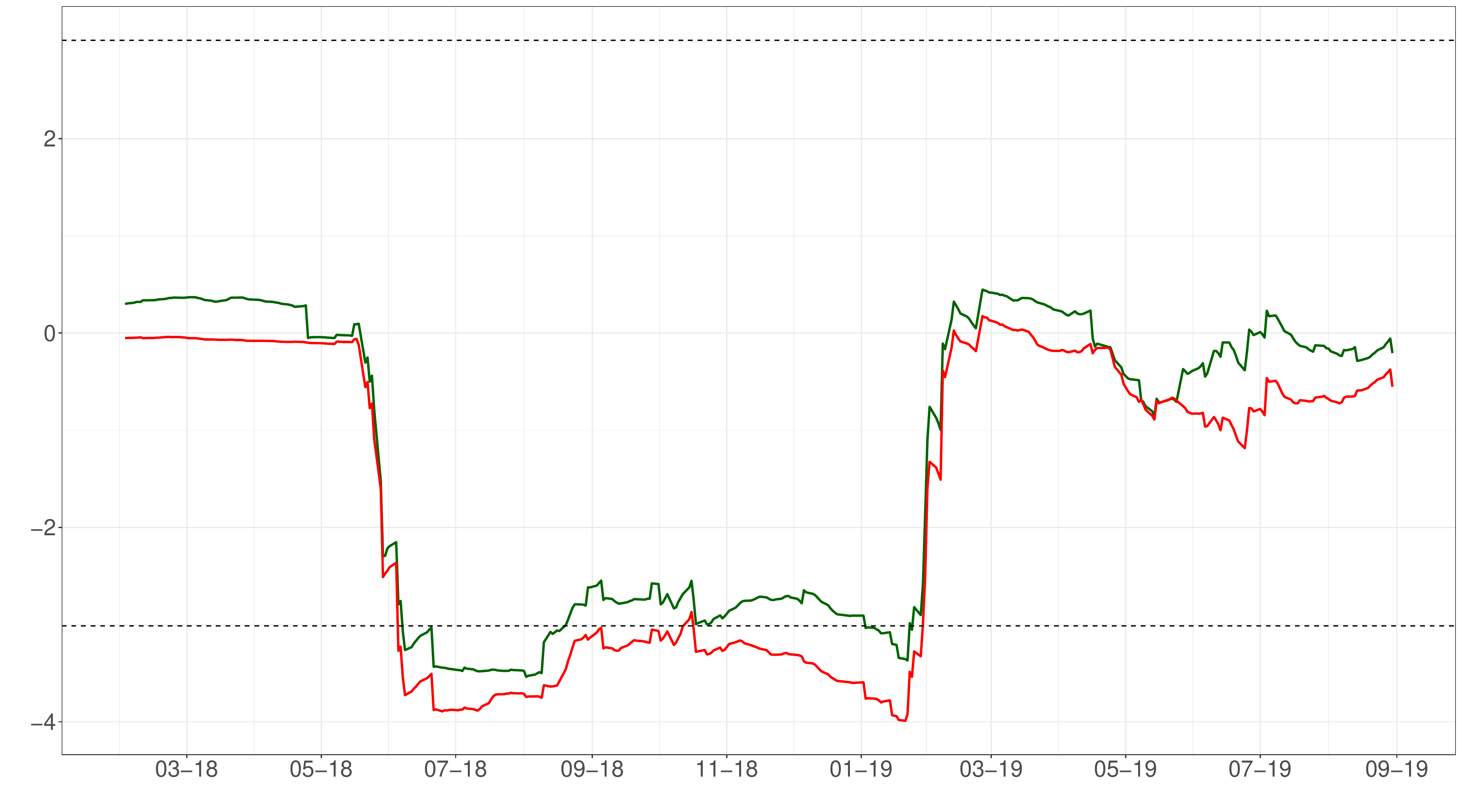}
\end{center}
\begin{center}
\includegraphics[scale=0.25]{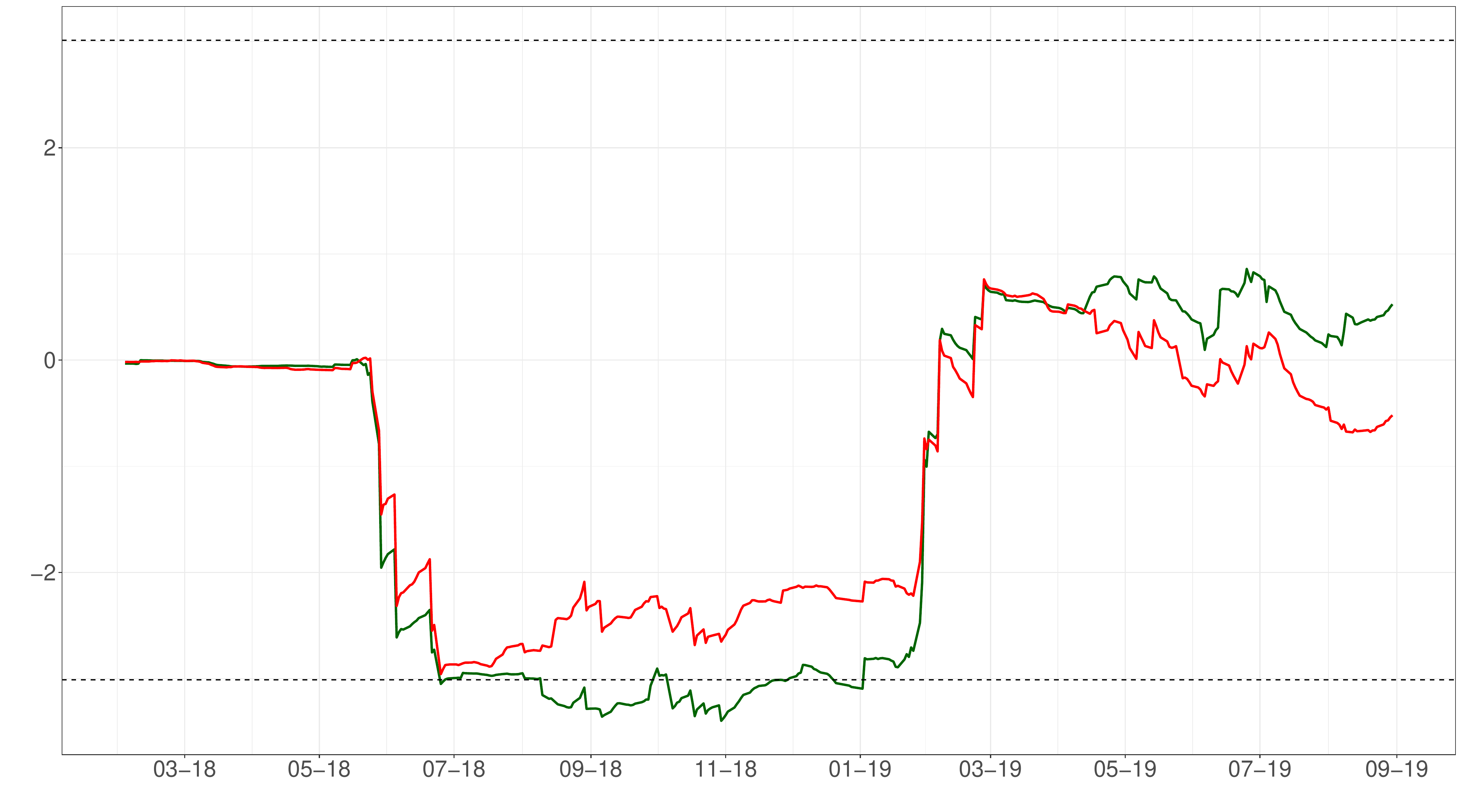}
\end{center}
{\footnotesize Notes: The Fluctuation test compares the forecasting performance of equation (\ref{eq1}): $\Delta Spread_{t + 1} = \alpha ^{q} + \delta_{0}^{q} \Delta Spread_{t}  + \delta_{1}^{q} X_{t} + \gamma^{q} LM_{t-h} + \beta^{q} Emotion_{t-h} + \epsilon_{t+1}$ against the one of the benchmark model $\Delta Spread_{t + 1} = \alpha ^{q} + \delta_{0}^{q} \Delta Spread_{t}  + \delta_{1}^{q} X_{t} + \gamma^{q} LM_{t-h} + \epsilon_{t+1}$ where $\beta^{q}$ is set to zero and with $q=0.95$. The red line is for $Emotion_{t}=Panic_{t}$, the green line is  $Emotion_{t}=Distress_{t}$. Negative values indicate that model in equation (\ref{eq1}) outperforms the benchmark model. The dashed line indicates the critical value of the Fluctuation test statistic at the 5 per cent significance level. When the estimated test statistic is below the negative critical value line, the model forecasts significantly better than the benchmark.}

\end{minipage}

\end{figure}

\begin{figure}[H]
\begin{minipage}{\textwidth}
    \caption{Fluctuation test for Spain (Domestic and international events) with $h=1$ (top) $h=5$ (bottom).}
    \label{fluctuationspain}
\begin{center}
\includegraphics[scale=0.25]{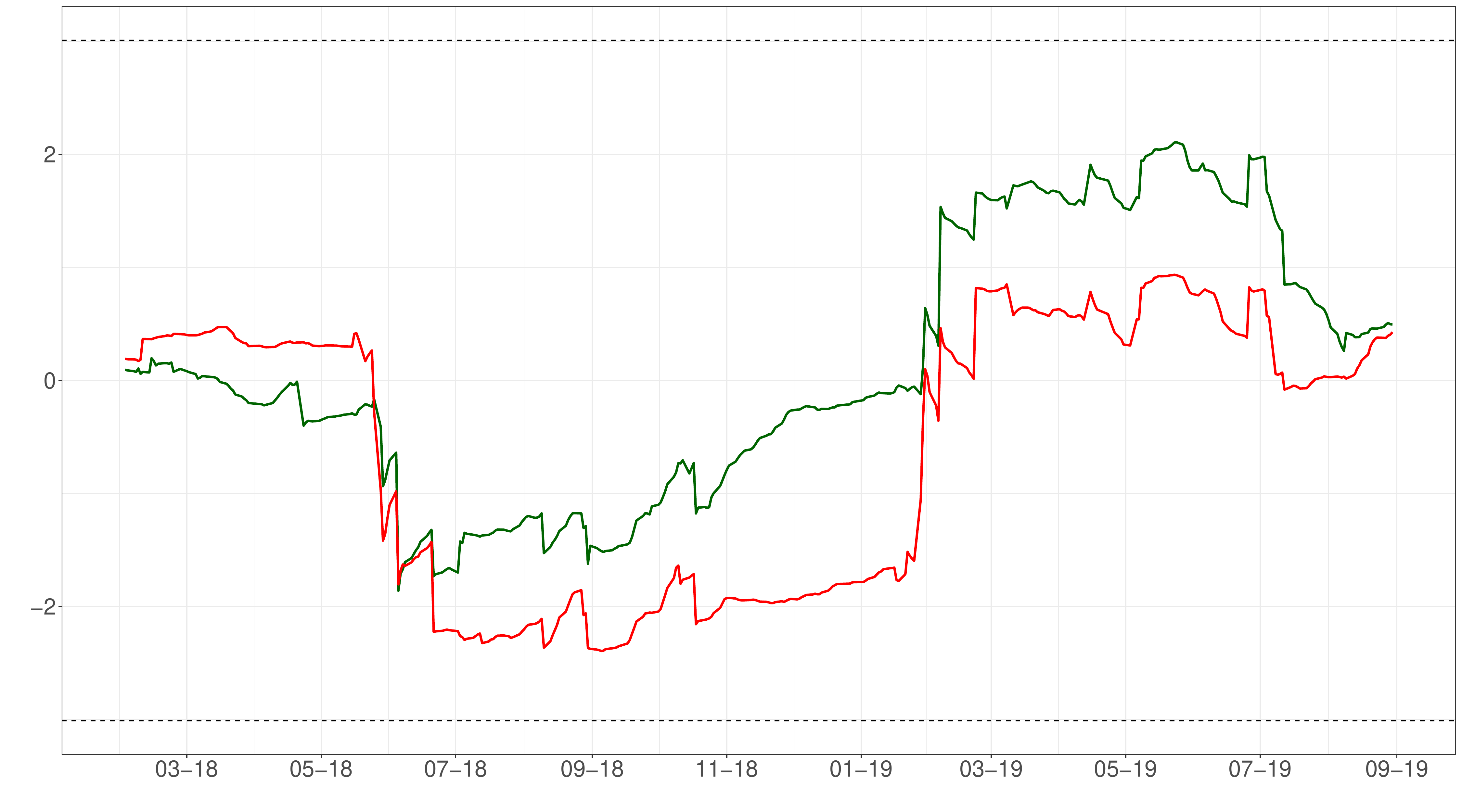}
\end{center}
\begin{center}
\includegraphics[scale=0.25]{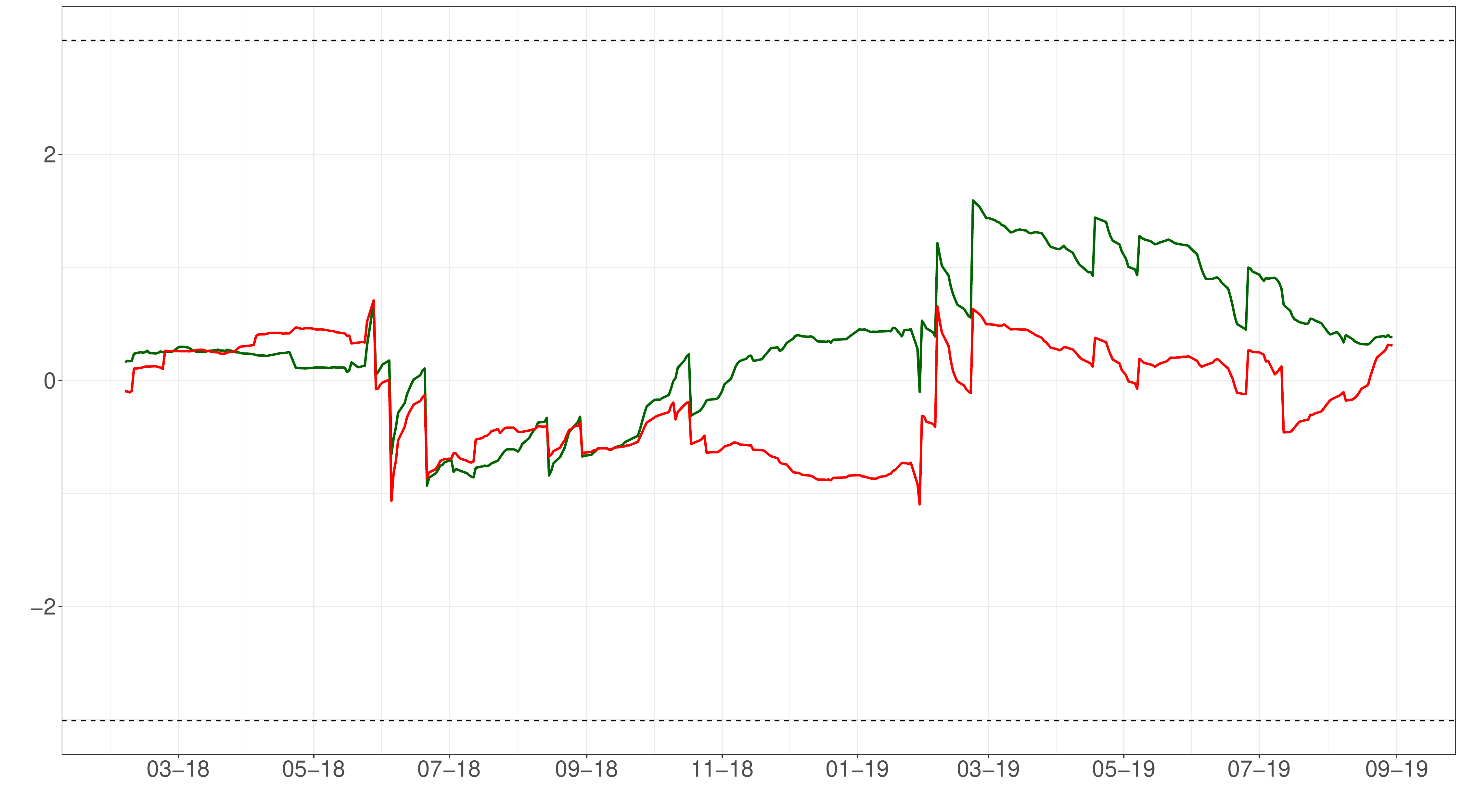}
\end{center}
{\footnotesize Notes: The Fluctuation test compares the forecasting performance of equation (\ref{eq1}): $\Delta Spread_{t + 1} = \alpha ^{q} + \delta_{0}^{q} \Delta Spread_{t}  + \delta_{1}^{q} X_{t} + \gamma^{q} LM_{t-h} + \beta^{q} Emotion_{t-h} + \epsilon_{t+1}$ against the one of the benchmark model $\Delta Spread_{t + 1} = \alpha ^{q} + \delta_{0}^{q} \Delta Spread_{t}  + \delta_{1}^{q} X_{t} + \gamma^{q} LM_{t-h} + \epsilon_{t+1}$ where $\beta^{q}$ is set to zero and with $q=0.95$. The red line is for $Emotion_{t}=Panic_{t}$, the green line is  $Emotion_{t}=Distress_{t}$. Negative values indicate that model in equation (\ref{eq1}) outperforms the benchmark model. The dashed line indicates the critical value of the Fluctuation test statistic at the 5 per cent significance level. When the estimated test statistic is below the negative critical value line, the model forecasts significantly better than the benchmark.}
\end{minipage}
\end{figure}

We now turn to the results of the \cite{Giacomini2010} Fluctuation test, reported in Figures \ref{fluctuationitaly} and \ref{fluctuationspain} for $h=1,5$. This test, by calculating the DM test over a rolling out-of sample window of a pre-specified size is more reliable than the simple DM statistics calculated by pooling observations from the full sample. The graphs show strong empirical evidence of time variation in the relative performance of the news-augmented models relative to the benchmark model that includes only financial variables. For Italy, and when setting $h=1$, the Fluctuation test for our emotion variables turns significant relative to the benchmark soon after at the start of the political crisis, with Panic turning significant on the $5$th of June, followed by Distress on the $8$th of June. 
For Panic we observe an increase in the significance of the test starting from late October 2018, when Moody's downgraded the Italian senior unsecured bond rating, until the end of January 2019.
As for Distress, it remains significant until mid-August 2018. 
In the case of the 5-steps ahead forecasts, only Distress falls below the critical value line, over the period starting in mid-August and ending in November 2018 with the exception of a few days at the beginning of October. We observe that this is a period of persistent concern for market participants given the negative outlook for the Italian economy provided by Fitch in September 2018 and the tension between Italy and the European Union concerning the Italian financial and economic budget plan later that year. 
Overall, these graphs confirm the better forecasting ability of Panic at a short time horizon ($h=1$) and of Distress at a longer time horizon ($h=5$). As for Spain, our emotions do not provide statistically significance neither at long nor at short time horizon. While distress and panic triggered by international events contribute to the in-sample fitting performance of our model for Spain, the predictive power of our emotions vanishes in the out-of-sample exercise.

\section{Conclusions}
\label{S6}

In this paper we studied the effect of news emotions to help predicting changes in government yield bond spread in Italy and Spain. Empirical results suggested that augmenting quantile regression with selected daily news emotions significantly improve the predictive power of conventional models for government yield bond spread, for both Italy and Spain, also after controlling for the overall negativity in the news text measured by the \citet*{Loughran_McDonald_2011} dictionary. One interesting finding is that the focus of news seems to play an important role in explaining variations in the government yield spreads, with Italy mostly worried about the domestic economic situation, and Spanish anxiety mostly triggered by international events. When using our news-based indicators for forecasting future variations of government yield bond spreads, our emotion variables showed some forecasting ability only for Italy. In particular, for this country the relatively lower intensity emotion of Distress was able to forecast fluctuations in the spread as long as a week after, while the relatively higher intensity emotion of Panic improved forecasts of one day ahead changes in the spread. \newline

Overall, the analysis of emotions extracted from news seems a promising area of research, particularly useful for capturing future intentions of agents in financial markets. One interesting extension of this work is to consider a wider set of emotions, both positive and negative, and use them to forecast both tails of the distribution of government yield spreads. Future work could also consider adopting machine learning techniques to look at additional features of articles and their non-linear effect on the dependent variable. 
This will be the object of a future study.

\newpage

\section*{Acknowledgments}

The authors would like to thank the colleagues of the Centre for Advanced Studies at the Joint Research Centre of the European Commission for helpful guidance and support during the development of this research work.\\
We would like to thank Sebastiano Manzan, Luca Barbaglia, Javier J. Perez, Barbara Rossi, Fabio Trojani, Leonardo Iania, Laurent Callot, Guillaume Chavillon as well as the participants at the JRC Workshop on Big Data and Economic Forecasting 2019, New Techniques and Technologies for Official Statistics 2019, the Smart Statistics for Smart Application Conference of the Italian Statistical Society in Milan, Mining Data for Financial Application Workshop in Wurzburg, CFE 2019 Conference in London for helpful comments and remarks.\\

\newpage\phantomsection%
\bibliographystyle{harv}

\newpage

\section*{\large{Figures}}

\begin{table}[H]
\caption{List of events covered by newspapers with dates and descriptions.}
\footnotesize
\centering
\begin{tabular}{cll}
 \\[-1.8ex]\hline 
  \hline \\[-1.8ex] 
  ID & Date & Event \\ 
  \\[-1.8ex]\hline 
  1 & 29 August 2015 & Greek government debt crisis \\ 
  2 & 24 February 2016 & EU-wide stress testing \\ 
  3 & 23 June 2016 & Brexit \\ 
  4 & 4 December 2016 & Italian constitutional referendum \\ 
  5 & 23 April 2017 & French political elections \\ 
  6 & 7 June 2017 & Rumours early Italian political elections \\ 
  7 & 6 September 2017 & Discussion on tapering \\ 
  8 & 1 October 2017 & Catalan referendum \\ 
  9 & 5 February 2018 & Stock market crash and increase in volatility\\ 
  10 & 4 March 2018 & Italian political elections \\ 
  11 & 29 May 2018 & Political crisis in Italy and in Spain\\ 
  12 & 6 September 2018 & Fitch confirms negative Italian outlook \\ 
  13 & 19 October 2018 & Moody's downgrade of the Italian senior unsecured bond ratings\\ 
  14 & 20 December 2018 & Italian agreement with Brussels on the budget deficit \\ 
  15 & 7 February 2019 & EU publishes Winter 2019 Economic Forecast \\ 
  16 & 26 May 2019 & European parliament elections \\ 
   \hline
\end{tabular}
\label{tab_events}
\end{table}

\newpage

\newpage

\newpage

\newpage

\newpage

\newpage

\newpage

\newpage

\newpage

\newpage

\newpage

\section*{\large{Appendix A}}

\label{Appendix_3A}


Below we report the list of outlets by country of origin that have been selected for the analysis.
\subsubsection*{Italy}

Il Sole 24 Ore, Borsa Italiana, Italia Oggi, Milano Finanza, Ansa, il Giornale, Finanza, Wall Street Italia, la Repubblica, Investire Oggi, Libero Quotidiano, il Messaggero Economia, il Fatto Quotidiano, il Corriere della Sera, La Stampa Finanza, Huffington Post Italy, La Stampa, trend-online.com, teleborsa, tradelink, il Tempo, finanza on-line, il Sussidiario.

\subsubsection*{Spain}

ABC, El Mundo, El Pais, El Economista, Cincodias, Expansion, Libre mercado, Finanzas, Economia Finanzas, Bolsa, Info Bolsa, El Comercio, El dia, El progreso, El confidencial, El Confidencial Digital, Huffington Post Spain, La razon, Negocios, El Diario.

\vspace{1cm}

 \newpage

\newpage

		
\section*{\large{Tables}}			
\appendix
\label{c:A_ITYC}

\renewcommand{\arraystretch}{0.80}
\begin{table}[H]
\begin{center}
    \caption{Descriptive statistics for the government yield spread in Italy and Spain.}  
      \begin{tabular}{clcccc|cccc}
  \hline
  \footnotesize
   & &  &  Italy  &  &  &  & Spain    &   \\ 
  \hline
  & Year & Mean & S.d. & 5th perc & 95th perc & Mean & S.d. & 5th perc & 95th perc \\ 
  \hline
  2015 & $Spread$ & 115.40 & 15.20 & 95.47 & 147.69 & 122.25 & 14.80 & 99.16 & 146.50\\ 
   & $\Delta Spread$  & -0.02 & 5.92 & -8.58 & 9.27  & 0.10 & 6.36 & -8.08 & 8.84 \\ 
   & $CRD$  & -0.01 & 1.60 & -2.68 & 2.55 & -0.06 & 1.37 & -2.32 & 2.27 \\ 
   & $LIQ$  & 0.00 & 0.14 & -0.20 & 0.20  & 0.00 & 0.27 & -0.40 & 0.40\\ 
   & $VSTOXX$  & 0.02 & 1.84 & -2.68 & 2.71  & - & - & - & -\\ 
   2016 & $Spread$ & 132.31 & 19.17 & 103.62 & 168.92 & 123.72 & 16.13 & 101.50 & 148.10\\ 
   & $\Delta Spread$  & 0.25 & 4.74 & -7.45 & 7.57 & 0.01 & 4.59 & -6.22 & 6.72\\ 
   & $CRD$ & -0.02 & 1.92 & -2.86 & 3.22 & 0.00 & 1.61 & -2.45 & 2.37\\ 
   & $LIQ$ & -0.00 & 0.11 & -0.13 & 0.20 & -0.00 & 0.14 & -0.20 & 0.20\\ 
   & $VSTOXX$  & -0.02 & 1.55 & -2.46 & 2.64   & - & - & - & -\\
   2017 & $Spread$ & 169.82 & 17.83 & 141.86 & 200.20 & 118.94 & 12.54 & 98.67 & 142.53\\ 
   &  $\Delta Spread$ & -0.01 & 4.25 & -6.94 & 7.07 & -0.02 & 3.73 & -5.96 & 5.56\\ 
   &  $CRD$ & 0.05 & 0.88 & -1.27 & 1.50 & 0.03 & 0.81 & -1.22 & 1.44\\ 
   &  $LIQ$ & 0.01 & 0.46 & -0.70 & 0.80 & 0.00 & 0.71 & -1.16 & 1.23\\ 
   & $VSTOXX$  & -0.02 & 1.09 & -1.28 & 1.53  & - & - & - & -\\ 
   2018 & $Spread$ &  213.59 & 68.38 & 122.86 & 308.33 & 96.71 & 17.89 & 68.56 & 122.46\\ 
   & $\Delta Spread$ & 0.36 & 9.38 & -13.51 & 15.10 & 0.02 & 4.08 & -5.12 & 6.08\\ 
   & $CRD$ & -0.06 & 1.15 & -1.92 & 1.77 & -0.06 & 0.87 & -1.57 & 1.21\\ 
   & $LIQ$ & -0.01 & 0.46 & -0.50 & 0.40 & -0.00 & 0.41 & -0.50 & 0.54\\ 
   & $VSTOXX$ & 0.04 & 1.81 & -2.13 & 2.59    & - & - & - & -\\
   2019 & $Spread$ & 243.91 & 28.94 & 189.04 & 279.42  & 94.97  & 16.87 & 70.79 & 118.57\\   
    & $\Delta Spread$ & -0.48 & 6.50 & -10.68 & 8.18 & -0.22 & 2.76 & -3.81 & 4.00 \\ 
    & $CRD$ & 0.09 & 0.99 & -1.75 & 1.59 & 0.02 & 0.79 & -1.35 & 1.30\\ 
    & $LIQ$  & 0.00 & 0.14 & -0.20 & 0.20 & 0.00 & 0.21 & -0.30 & 0.30 \\ 
    & $VSTOXX$ & -0.04 & 1.06 & -1.61 & 2.1  & - & - & - & -\\ 
   \hline
\end{tabular}
        \label{tab1}
    \end{center}

{\footnotesize Notes: Since $VSTOXX$ is calculated at European level, we only report this statistics for Italy as it is identical for Spain.}        
    \end{table}

\newpage

\end{document}